\PassOptionsToPackage{x11names}{xcolor}
\documentclass[
reprint,
superscriptaddress,
amsmath,amssymb,
aps,
prx,
longbibliography
]{revtex4-2}

\usepackage{graphicx}
\usepackage{dcolumn}
\usepackage{bm}
\usepackage[caption=false]{subfig}
\usepackage{placeins}
\usepackage{overpic}
\usepackage{natbib}
\usepackage{xcolor}
\usepackage{algorithm}
\usepackage{algpseudocode}
\usepackage{booktabs}
\usepackage[T1]{fontenc}
\usepackage{lmodern}
\usepackage{hyperref}
\hypersetup{
	colorlinks=true,
	linkcolor=blue,
	filecolor=blue,
	urlcolor=blue,
	citecolor=blue,
}
\DeclareMathOperator*{\argmin}{argmin}

\newcounter{algorithmAPS}
\renewcommand{\thealgorithmAPS}{\arabic{algorithmAPS}}

\newenvironment{apsalgorithm}[2]{
	\refstepcounter{algorithmAPS}
	\label{#1}
	\small
	\noindent{ALGORITHM \thealgorithmAPS.} #2\par
	\vspace{1.2ex}
	\hrule height 0.6pt
	\vspace{0.4ex}
	\hrule height 0.6pt
	\vspace{1ex}
	\begin{algorithmic}[1]
	}{
	\end{algorithmic}
	\vspace{1ex}
	\hrule height 0.6pt
	\vspace{0.4ex}
	\hrule height 0.6pt
}

\newcounter{apstable}
\renewcommand{\theapstable}{\Roman{apstable}}

\newenvironment{apstable}[2]{
	\begin{table}[t]
		\refstepcounter{apstable}
		\label{#1}
		\small
		
		\noindent{TABLE \theapstable.} #2\par
		\vspace{0.77ex}
		\hrule height 0.6pt
		\vspace{0.4ex}
		\hrule height 0.6pt
		\vspace{1ex}
		\centering
	}{
		\vspace{1ex}
		\hrule height 0.6pt
		\vspace{0.4ex}
		\hrule height 0.6pt
	\end{table}
}

\begin{document}
	\title{Controlling Defects and Probing Dynamics in Active Nematics with Deep Reinforcement Learning}

	\author{Russ Islam}
	\affiliation{Department of Physics, The University of Tokyo, 7-3-1 Hongo, Bunkyo-ku, Tokyo 113-0033, Japan}

	\author{Kyogo Kawaguchi}
	\affiliation{Department of Physics, The University of Tokyo, 7-3-1 Hongo, Bunkyo-ku, Tokyo 113-0033, Japan}
	\affiliation{Institute for Physics of Intelligence, The University of Tokyo, 7-3-1 Hongo, Bunkyo-ku, Tokyo 113-0033, Japan}
	\affiliation{Nonequilibrium Physics of Living Matter Laboratory, RIKEN Pioneering Research Institute, 2-2-3 Minatojima-minamimachi, Chuo-ku, Kobe 650-0047, Japan}
	\affiliation{Universal Biology Institute, The University of Tokyo, 7-3-1 Hongo, Bunkyo-ku, Tokyo 113-0033, Japan}

	\author{Yuto Ashida}
	\affiliation{Department of Physics, The University of Tokyo, 7-3-1 Hongo, Bunkyo-ku, Tokyo 113-0033, Japan}
	\affiliation{Institute for Physics of Intelligence, The University of Tokyo, 7-3-1 Hongo, Bunkyo-ku, Tokyo 113-0033, Japan}

	\date{\today}

	\begin{abstract}
		Topological defects govern much of the flow behavior and orientational order in active nematics, making their control relevant for active matter physics, smart materials, and microfluidics. Applied activity patterns can induce self-propulsion of active nematic defects, but general-purpose methods for exploiting this effect to control defects remain largely unexplored. Here we use deep reinforcement learning (RL) to perform minimum-time position control of $+1/2$ defects in hybrid lattice Boltzmann simulations of active nematodynamics. Spatiotemporally patterned activity, implemented as a control field in the active stress, steers defects through microchannel geometries and reveals finite-time reachable regions of defect position space. Reachability is shaped by director anisotropy, homeotropic wall anchoring, and the allowed activity patterns: local patterns steer defects in free domains but fail in junctions, whereas global patterns open otherwise inaccessible channels. In constrained geometries, the original defect may be unable to reach some goals intact, but controlled pair creation enlarges the effective reachable set by transferring control to a newly created $+1/2$ defect. The trained RL controllers outperform static and rule-based baselines, and controllers trained only on simple junctions can be combined without fine-tuning into a meta-controller that successfully steers defects through a larger test maze. Free energy visualizations show that guided defects write persistent, history-dependent distortions into the director field that can later be partially erased by $-1/2$ defects. Thus, RL-based control uncovers how confinement, anchoring, actuation geometry, and defect creation determine reachable motion in active nematics, providing a framework for other control tasks in soft and active matter.
	\end{abstract}

	\maketitle

	\section{Introduction}
	\label{sec:intro}

	\subsection{Background and Motivation}
	\label{subsec:background}

	Active matter systems have drawn much attention for their varied nonequilibrium behavior, attracting interest from nonequilibrium statistical physics, materials science, and biophysics. Because they break energy and momentum conservation and lack detailed balance, active materials cannot be described using only standard equilibrium tools. Biological matter is especially far from equilibrium: chemical and biomechanical regulation sustains life in regimes where passive materials would simply relax from one equilibrium configuration to another \cite{Takatori_2025}.

	Feedback control lies at the heart of bioregulation at every scale \cite{Alvarado_2026}. Although our knowledge of biochemical feedback loops has matured considerably, our understanding of the interplay between control and biophysics remains limited \cite{Hana_biofeedback_2021}. Cellular transport, cell motility, self-assembly, and jamming suppression may all benefit from a better grasp of active matter control \cite{Doostmohammadi_2018, Yao_jamming_2025}. Flexible control frameworks would also give experimentalists a way to probe model systems efficiently and reproducibly.

	Successful control of active matter would also have technological value. Smart materials, microfluidics, and microrobotics all require reliable ways to actuate and steer nonequilibrium materials. Existing microfluidic devices largely rely on confined isotropic fluids driven by miniature pumps and valves, but active fluids, especially active nematics, could enable devices with more advanced operations \cite{Zhang_Catchmark_2011, Kos_2019, Kos_Ravnik_2020}. By augmenting or replacing functions now performed by mechanical components, active fluids could convert microscale energy input into macroscale mechanical response and useful work. Micromixing, colloidal assembly, and programmable matter would all benefit from reliable active matter control \cite{manz_2021, Zhang_Mozaffari_de_Pablo_2022, nematic_bits, nematic_microrobots, Yang_Liu_2025}.

	Active nematics are central both to fundamental theory and to applications in biophysics and engineering. Soft matter with nematic order appears throughout biology, from the cytoskeleton to bacterial colonies to morphogenesis \cite{Kawaguchi_Kageyama_Sano_2017, Ravichandran_2025}. Active nematic fluids exhibit phenomena such as active turbulence and wetting, but topological defects are especially important because they organize flow and orientational structure.

	In extensile active nematics, increasing the activity near a $+1/2$ defect drives self-propulsion along the direction of its comet-like head, and this defect motion reshapes the surrounding flow and director fields \cite{Doostmohammadi_2018, Ramaswamy_2010, RevModPhys.85.1143, Giomi_2015, Thampi_Golestanian_Yeomans_2014, Kos_Aplinc_Mur_Ravnik_2019}. Spatially patterned activity, for example in the form of applied light, can therefore serve as a physical control field for defect motion. Some experiments have already leveraged light patterns to shape active nematic flows \cite{Ross_Phillips_2019, Zhang_Redford_2021, Lemma_Dogic_2023, Zarei_Dogic_2023}, and simulations with manually designed static patterns have guided defects and performed logic operations analogous to transistor gates \cite{Zhang_Mozaffari_de_Pablo_2022}. These developments motivate a basic physical question: given a confined geometry, boundary anchoring, and a restricted set of activity patterns, which defect motions are achievable in finite time?

	Despite the importance of feedback control in active and biological matter, broadly adaptable controllers remain difficult to construct. High entropies, nonlocal interactions, and complicated stress responses make soft and active materials difficult to model, let alone control. These materials are usually described by systems of PDEs that may respond to applied active stress in complex, chaotic ways, often making classical control techniques ineffective or impractical.

	Even the transport of one defect through an active nematic medium presents many difficulties: fluid anisotropy complicates defect mobility, long-range correlations couple the defect to distant distortions, and the possible activity profiles form a high-dimensional decision space. Progress in active matter control therefore requires methods that can capture relationships among many variables while remaining flexible across different control tasks.

	Reinforcement learning (RL) is a natural candidate for these control problems. Unlike supervised learning, RL does not require labeled input-output pairs; instead, it learns from rewards assigned to actions \cite{Sutton_Barto_2020, Lapan_2024}. During training, an agent receives observations from an environment, selects an action, applies that action, and receives a reward. The agent's observation-to-action mapping is periodically updated to maximize expected future rewards, so the agent learns by trial and error through its own interactions with the environment.

	This learning structure is well-suited to flow fields and other high-dimensional states, where gathering and labeling data in advance is impractical \cite{Mnih_2015, Rabault_2019}. In particular, deep RL has already been used for demanding control tasks in the physical sciences, including quantum control \cite{Bukov_2018, An_Zhou_2019, Wang_Ashida_Ueda_2020, Baba_Yoshioka_Ashida_Sagawa_2023, Niu_Boixo_Smelyanskiy_Neven_2019} and fluid dynamics \cite{Rabault_2019, Novati_2021, Font_2025}.

	Meanwhile, existing active matter control studies have largely relied on classical control theory, including work on chirality switching \cite{Norton_2020}, active drop transport \cite{Shankar_Raju_Mahadevan_2022}, and minimum-dissipation control of active particles \cite{Davis_2024}. Designed rulesets and manually constructed activity patterns have also been used for active nematic defect control \cite{Shankar_Scharrer_Bowick_Marchetti_2024, Zhang_Mozaffari_de_Pablo_2022}, and classical feedback control of active nematic flow speeds has recently been realized in experiment \cite{Nishiyama_Dogic_2025}.

	The application of reinforcement learning to active matter remains more limited. The use of RL for controlling microswimmers and self-propelled disks has been explored to some extent \cite{Muinos_2021, Falk_Alizadehyazdi_Jaeger_Murugan_2021, Bulusu_Zottl_2025}. RL has even been applied to active nematic and active polar systems, but these studies have focused on reducing the total number of defects \cite{Hou_2025}, tailoring local interactions between paired defects \cite{Floyd_Dinner_Vaikuntanathan_2025}, or rotating an isolated polar defect inside a small circular domain \cite{Singh_2025}. Arbitrary defect control has not yet been accomplished, especially for larger and more complicated domains. Deep RL may therefore provide a route to general-purpose defect controllers while also identifying the physical constraints that govern active nematic control landscapes.

	\begin{figure*}[t]
	\centering
	\includegraphics[width=\textwidth]{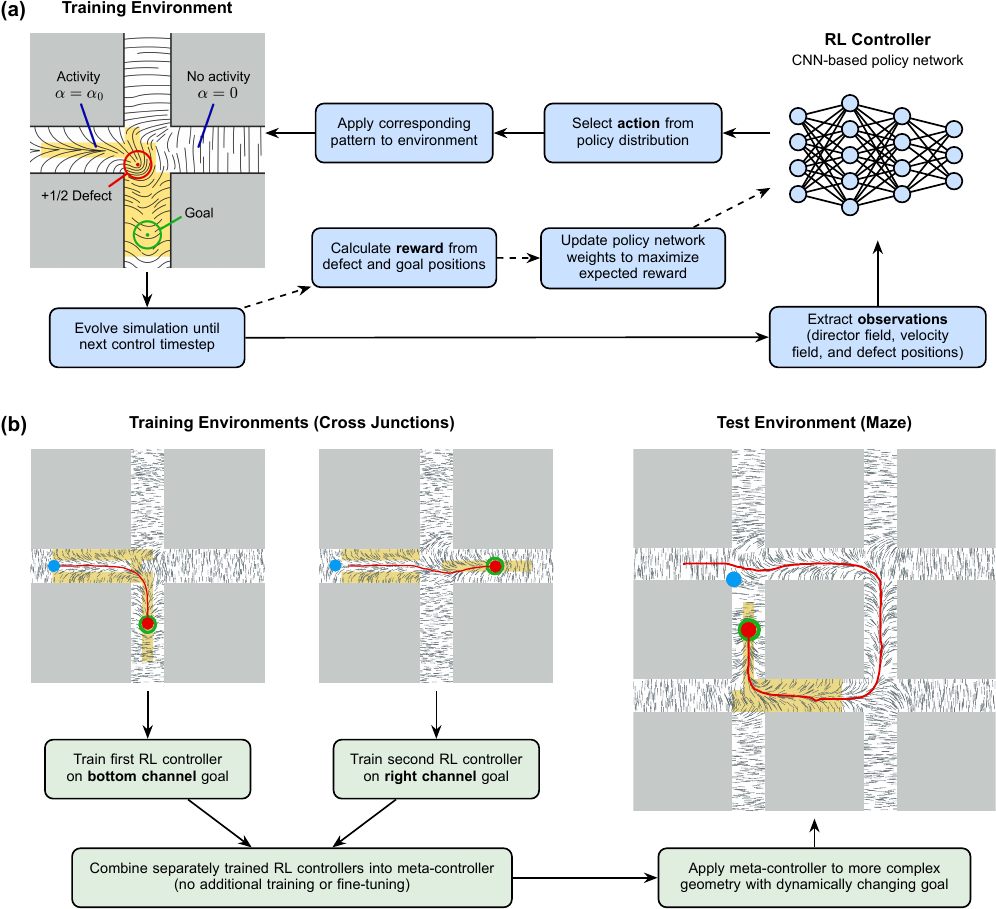}
	\caption{Overview of the method and main results. (a) We train deep RL controllers by applying activity patterns (actions) to active nematic defects in hybrid LBM simulations. Rewards depend on the distance between the defect and the goal position. This encourages the controller to steer the defect to the goal in minimum time. The controller closes the loop by using observations from the environment to choose the next action. (b) RL controllers trained on separate environments may be combined into one meta-controller, without additional training, and applied to new test geometries. In our test maze, this meta-controller successfully steers a defect through a sequence of waypoints. These trajectories reveal insights into reachability, free energy evolution, and memory effects in constrained microchannel geometries.}
	\label{fig:paper_key_result}
	\end{figure*}

	\subsection{Overview of Key Results}
	\label{subsec:overview}

	\begin{figure*}[t]
	\centering
	\includegraphics[width=\textwidth]{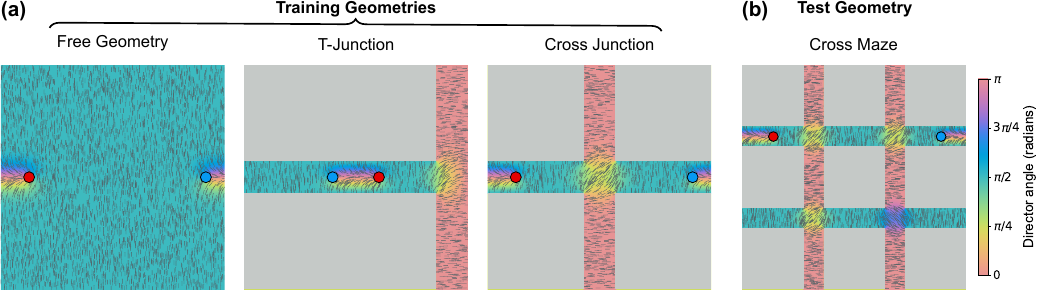}
	\caption{Initial director fields for the domain geometries considered in this work, shown with cyclic colormaps. $+1/2$ ($-1/2$) defects are shown as red (blue) circles. Obstacles are shown in solid gray. (a) Training geometries consist of the free geometry, the T-junction, and the cross junction. All have lattice dimensions of 420-by-420 sites. (b) The test geometry is a 660-by-660-site maze made of four interconnected cross junctions. All channels are 60 lattice sites wide. Note the broken symmetry of the director fields within all junctions, which strongly affects defect mobility.}
	\label{fig:initial_fields}
	\end{figure*}

	To our knowledge, we are the first to use deep reinforcement learning to develop feedback controllers for steering individual defects through microchannels. We outline our method and main results in Fig.~\ref{fig:paper_key_result}. The central objective is minimum-time position control of a $+1/2$ defect, but the learned activity sequences also expose the structure of the underlying active nematic control problem. The main physical result is that confined defect motion is governed by finite-time reachability landscapes: the accessible region depends on director anisotropy, channel geometry, and the admissible activity patterns.

	The deep RL training loop is shown in Fig.~\ref{fig:paper_key_result}(a). The controller selects an activity pattern from a set of allowable patterns and applies it to a microchannel-confined active nematic simulation based on the hybrid lattice Boltzmann method (LBM). After the simulation evolves to the next control step, velocity and director field values are sent back to the controller as observations. This feedback is used to choose the next action, and a reward computed from the defect-to-goal distance is used to update the controller's neural network until training converges.

	Our reachability analyses reveal several physical constraints on defect mobility. Local activity patterns can steer defects in free domains, but they are unable to push defects through channel junctions with homeotropic walls. Global activity patterns can open transport pathways through intersections, though some turns remain inaccessible to the original defect. In those cases, controlled pair creation can change the effective reachable set by transferring the control objective to a newly created $+1/2$ defect.

	Quantitative comparisons show that the trained controllers outperform naive static patterns and our own rule-based controller, especially when successful control requires suppressing or exploiting defect pair creation. As demonstrated in the test maze geometry (Fig.~\ref{fig:paper_key_result}(b)), controllers trained on simple geometries can also be combined without further training into a meta-controller that can steer a defect through a larger network of junctions. The same trajectories reveal persistent changes in the Landau-de Gennes free energy landscape, showing that defect control can write and partially erase memory in the director field.

	We proceed by first defining the active nematic control problem, then using reachable sets to characterize the accessible defect-motion landscape, and finally asking whether deep RL can find minimum-time control policies within that landscape and apply them to a larger, more complex geometry.

	\section{Methodology}
	\label{sec:method}

	Our controller development procedure has three components. First, we simulate coarse-grained active nematics using the hybrid lattice Boltzmann method. Second, we perform reachability analyses on different geometries and control pattern sets to characterize the control landscape. Third, we train defect controllers via deep reinforcement learning. The same procedure can be extended to other geometries and to more complex soft matter control tasks.

	\subsection{Active Nematic Model and Simulation Method}
	\label{subsec:physical_model}

	The physical system is a two-dimensional active nematic described by the symmetric, traceless tensor order parameter $Q(\textbf{x}, t)$ and the velocity field $u(\textbf{x}, t)$. The director dynamics obey the Beris-Edwards equation \cite{Beris_Edwards_1994}
	\begin{equation}
		\frac{\partial Q}{\partial t} + u \cdot \nabla Q - S = \Gamma H,
		\label{eq:main_beris_edwards}
	\end{equation}
	where $S$ is an advection term, $\Gamma$ is the molecular field strength, and $H=-\delta f/\delta Q$. Here $H$ is the molecular field derived from the Landau-de Gennes free energy density
	\begin{equation}
		f = \frac{A}{2} \text{Tr} \left( Q^2 \right) + \frac{B}{3} \text{Tr} \left( Q^3 \right) + \frac{C}{4} \left( \text{Tr} \left( Q^2 \right) \right)^2 + \frac{L}{2} \left( \nabla Q \right)^2 .
		\label{eq:main_ldg_free_energy}
	\end{equation}
	The velocity field is evolved using a damped Navier-Stokes equation,
	\begin{equation}
		\rho \left( \frac{\partial}{\partial t} + u \cdot \nabla \right) u = \nabla \cdot \Pi - \mu u,
		\label{eq:main_navier_stokes}
	\end{equation}
	with total stress $\Pi=\Pi^p+\Pi^a$. The control enters through the active stress
	\begin{equation}
		\Pi^a(\textbf{x}, t) = -\alpha(\textbf{x}, t) \, Q(\textbf{x}, t),
		\label{eq:main_active_stress}
	\end{equation}
	where $\alpha>0$ corresponds to an extensile active nematic in our convention. Full expressions for $S$ and the passive stress $\Pi^p$ are given in Appendix~\ref{sec:app_be_eq}.

	At each control step, the controller chooses an activity pattern $P_j(\textbf{x})$ from a specified control pattern set. In the simulations reported here, these patterns are binary: activated regions have $P_j=1$, and the rest of the fluid has $P_j=0$. The activity field is then
	\begin{equation}
		\alpha(\textbf{x}, t) = \alpha_0 P_j(\textbf{x})
	\end{equation}
	for the duration of one control interval $\tau$, after which the controller selects a new pattern. Here $\alpha_0$ is the fixed activity level applied in activated regions (see Table~\ref{tab:lbm-params} in Appendix~\ref{sec:app_be_eq} for its numerical value). Physically, activating a region locally increases the extensile active stress. Near a $+1/2$ defect, this stress drives motion along the defect head, while the induced flow and director distortions couple the local actuation to the rest of the channel geometry.

	\begin{figure*}[t]
	\centering
	\includegraphics[width=\textwidth]{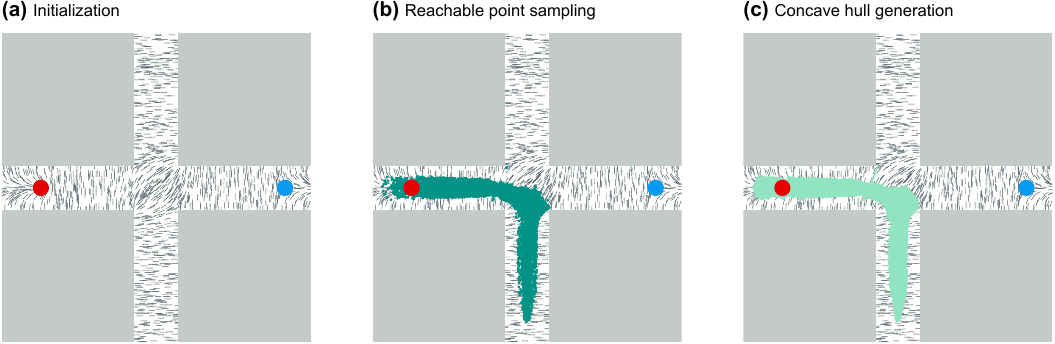}
	\caption{Steps involved in estimating the reachable set of defect positions for a given domain geometry and control pattern set. (a) An initial configuration is specified, including director field values and initial defect positions. (b) Random actions are sampled from the control pattern set and applied to the $+1/2$ defect over many control steps. The simulation is reset whenever the maximum episode length is reached. All positions reached by the $+1/2$ defect (dark green points) are recorded. (c) A concave hull (light green) is generated from the reached positions. This hull gives a polygonal estimate of the reachable set.}
	\label{fig:reach_set_est_method}
	\end{figure*}

	Widely used in nematics research, the hybrid lattice Boltzmann method is the foundation of our simulations \cite{yeomans_hlbm, Zhang_LBM_2016}. We model active nematics using coarse-grained fields for the local mean velocity and director orientation. The lattice Boltzmann method computes the velocity field, while finite differences compute the director field, thereby solving the active Beris-Edwards model.

	The hybrid LBM is flexible: it can handle various boundary conditions, complex geometries, multiphase flows, and immersed bodies such as colloids, making it well-suited for modeling flowing matter in geometries relevant to microfluidics and biophysics \cite{Kruger_Kusumaatmaja_Kuzmin_Shardt_Silva_Viggen_2017, Mohamad_2019, Succi_2022}. It is also highly parallelizable. Since RL controller training would be nearly impractical without GPU acceleration, for the present work we implemented an optimized version of the hybrid LBM in C and CUDA (Appendix~\ref{sec:app_be_eq}).

	One RL training episode consists of $2 \times 10^5$ simulation steps. This takes roughly 12 hours of wall time on a single CPU core, or 4 hours on a modern multi-core CPU. With CUDA acceleration, one episode completes in about 30 seconds on a single consumer-grade GPU. This speedup of two to three orders of magnitude makes hybrid-LBM-based deep RL training feasible, even when convergence requires $10^5$ training steps or more.

	We consider three representative domain geometries as training environments, with initial director fields and defect positions shown in Fig.~\ref{fig:initial_fields}. The free geometry contains no walls or obstacles, has a vertically oriented director field, and uses periodic boundaries on all four sides. As the simplest environment, it serves as a benchmark.

	The T-junction geometry consists of a central horizontal channel and two shorter vertical channels, with open (free) boundary conditions enforced at all three outlets. On the other hand, the cross junction contains four channels that meet in a central intersection, with periodic boundaries at all outlets. These latter two geometries are typical building blocks of larger microfluidic circuits. In all wall-confined geometries, no-slip velocity boundary conditions and infinite homeotropic (wall-normal) director anchoring are enforced along obstacle walls.

	Note that the T-junction and cross junction are initialized with broken symmetry: the director fields inside the intersections point 45 degrees from southwest to northeast. Following Zhang et al. \cite{Zhang_Mozaffari_de_Pablo_2022}, these initial director fields produce defect-free intersections, and we will soon show that they strongly bias subsequent defect mobility.

	These geometries span different wall and channel configurations, outlet boundary conditions, and initial defect positions, and therefore present different control challenges. Since results for the T-junction and cross junction are similar, we focus on the free geometry and the cross junction in the main text, with T-junction results shown in Appendix~\ref{sec:app_additional_results}. In Section \ref{sec:controller_perf}, we also consider a fourth geometry, the test maze of Fig.~\ref{fig:initial_fields}(b), to assess the generality of the learned RL policies.
	
	Our model and control fields are idealized, but they are motivated by experimentally accessible forms of patterned actuation. The most direct experimental setting is a photoactivatable microtubule-kinesin active nematic, where light patterns can modulate local active stress \cite{Ross_Phillips_2019, Zhang_Redford_2021, Lemma_Dogic_2023, Zarei_Dogic_2023}. Epithelial sheets and other cell-sheet nematics provide a complementary biological context, since defect positions influence collective dynamics and morphogenesis \cite{Saw_Doostmohammadi_2017, Vafa_Mahadevan_2022, Bera_Notbohm_2025}. In both settings, the binary activity patterns used here should be viewed as a controlled starting point for studying how finite-resolution, noisy, or delayed actuation affects defect steering.

	\subsection{Reachability Analysis}
	\label{subsec:reachability_analysis}

	Next, we conduct a reachability analysis for each domain geometry to quantify and visualize the mobility of the $+1/2$ defect under specified constraints. The initial defect position, domain geometry, and control pattern set determine which positions the defect can reach within a fixed number of control steps; the set of all such positions is the \emph{reachable set}.

	The goal is to sample and characterize the reachable set for a given environment and action space. These analyses reveal which target states are feasible or infeasible, distinguish easy control tasks from difficult but achievable ones, and expose how active nematic dynamics shape the control landscape \cite{Kirk_2012, lew_2022}. Thus, the reachable set is not only a control-theoretic tool; it is an intuitive physical picture of how the active material, boundaries, and allowed activity patterns convert local active stress into defect motion.

	Fig.~\ref{fig:reach_set_est_method} summarizes our reachable set estimation algorithm, and Appendix~\ref{sec:app_algos} gives the corresponding pseudocode. The first stage specifies four parameters: (1) the domain geometry, (2) the initial velocity and director fields, and hence the initial $+1/2$ defect position, (3) the control pattern set, and (4) the number of control steps used to evolve the system. In control theory, the control pattern set is termed the \emph{set of admissible controls} \cite{Kirk_2012}. The reachable set depends strongly on these controls, so distinct admissible control sets can produce different reachable sets even in the same geometry.

	In the second stage, we apply a randomly selected activity pattern at each control step. One control step occurs every $10^4$ simulation timesteps, which we denote by $\tau$, and between control steps the chosen activity pattern is held fixed as the simulation evolves. The current defect position is recorded at each control step. When the maximum episode length is reached, the system is reinitialized and another episode begins; we also reset if the defect is inadvertently annihilated. In our analyses, the maximum episode length is 20 control steps, and we repeat this process for $10^4$ control steps, or roughly 500 episodes. This yields a dense collection of random samples from the reachable set.

	By itself, this collection of positions is an unwieldy representation of the true reachable set. In the final stage, we compute the concave hull, or alpha shape, of the sampled points \cite{lew_2022, nandy_1994, daniels_1997, concave_hull}. This hull approximates the reachable set boundary with a polygon. Since the sampled reachable set is a subset of the true reachable set, the generated hull is usually a subset as well. Given a geometry and goal position, we perform reachability analyses on different control pattern sets until a hull contains the goal. For the control tasks studied below, this analysis also guides the choice of pattern set used for RL training.

	\subsection{RL Controller Training}
	\label{subsec:rl_training}

	Having estimated which targets are physically reachable for each admissible control set, we next train feedback policies to find fast trajectories within those reachable regions. We use standard reward-shaping techniques to improve convergence; Appendix~\ref{sec:app_rl_params} lists these techniques, RL hyperparameters, and other implementation details. All controllers use the same value of $\tau$ as the reachability analyses.

	Our controllers use proximal policy optimization (PPO), a widely used policy-based RL method \cite{schulman2017ppo, Lapan_2024, stable-baselines3}. The agent contains a deep neural network that takes in observations and outputs a probability distribution over the action space. We use PPO for its flexibility and good convergence behavior; it is especially powerful when paired with a convolutional neural network (CNN). The CNN extracts features from continuous matrix-valued velocity and orientation data produced by our hybrid LBM simulations, compressing raw observations into encodings for use by the deep policy network.

	We compare the RL controllers against two baselines. The first is a static controller that applies the same pattern over the entire episode, providing an intuitive baseline similar to the static patterns used in \cite{Zhang_Mozaffari_de_Pablo_2022}. The second is an original rule-based controller that uses a simple greedy strategy to choose an action at each control step. In principle, the rule-based controller supports any discrete action space. For simplicity and interpretability, we use only local patterns, i.e., small defect-centered rectangles. The controller assumes that the defect will move from the near end of an activity strip to the far end, and then chooses the candidate pattern that brings the defect closest to the goal at the next step.

	Appendix~\ref{sec:app_algos} gives the associated pseudocode. The full code for our methodology, including trained model weights, is openly available at the GitHub repository \cite{github_rl_defect_control_code}. Supporting media files are available at \cite{github_rl_defect_control_media}.

	\section{Reachability Analysis Results}
	\label{sec:reachability_analysis_results}

	\begin{figure*}[t]
	\centering
	\includegraphics[width=\textwidth]{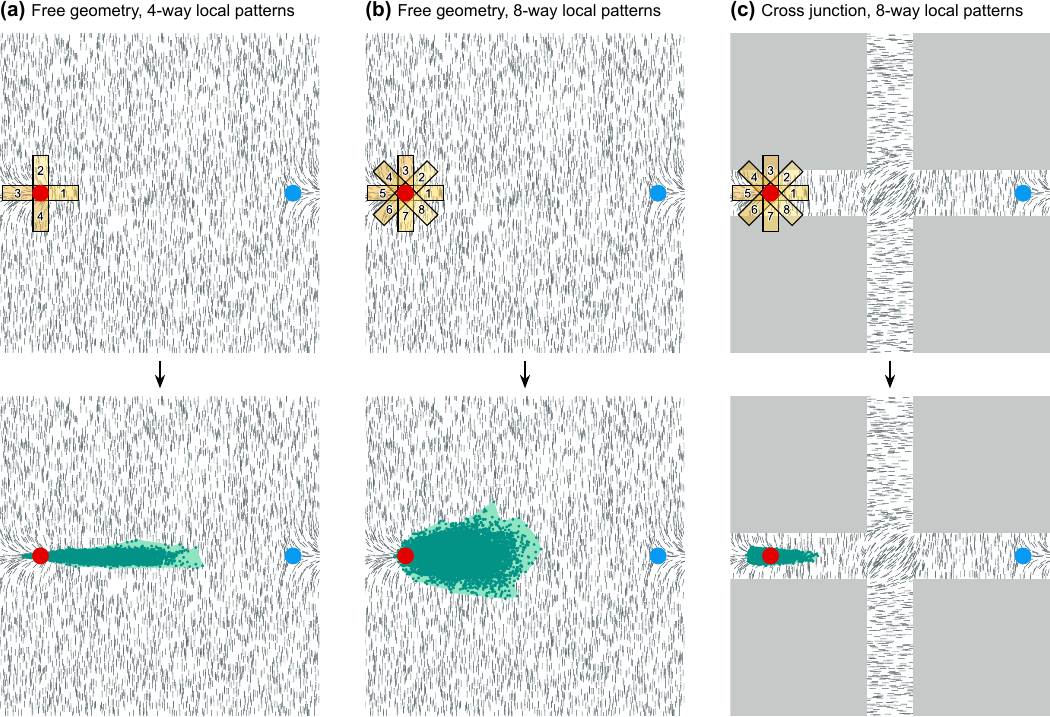}
	\caption{Reachability analysis with local control pattern sets. The top row shows initial configurations along with all available patterns in the set, while the bottom row displays corresponding estimates of the reachable sets. Pattern choices are numbered and filled with different shades of yellow for clarity. The episode length is 20 control steps. (a) For the 4-way local pattern set on the free geometry, the reachable set spreads quite far horizontally, but its vertical extent is severely limited. (b) For the 8-way local set on the free geometry, defect mobility in the vertical direction is greatly improved in comparison to the 4-way set. (c) For the 8-way local set on the cross junction, the estimated reachable set is restricted entirely to the left channel, covering only a small fraction of the simulation domain. Local patterns fail to guide the defect into other channels.}
	\label{fig:reach_sets_local}
	\end{figure*}

	\begin{figure*}[t]
	\centering
	\includegraphics[width=\textwidth]{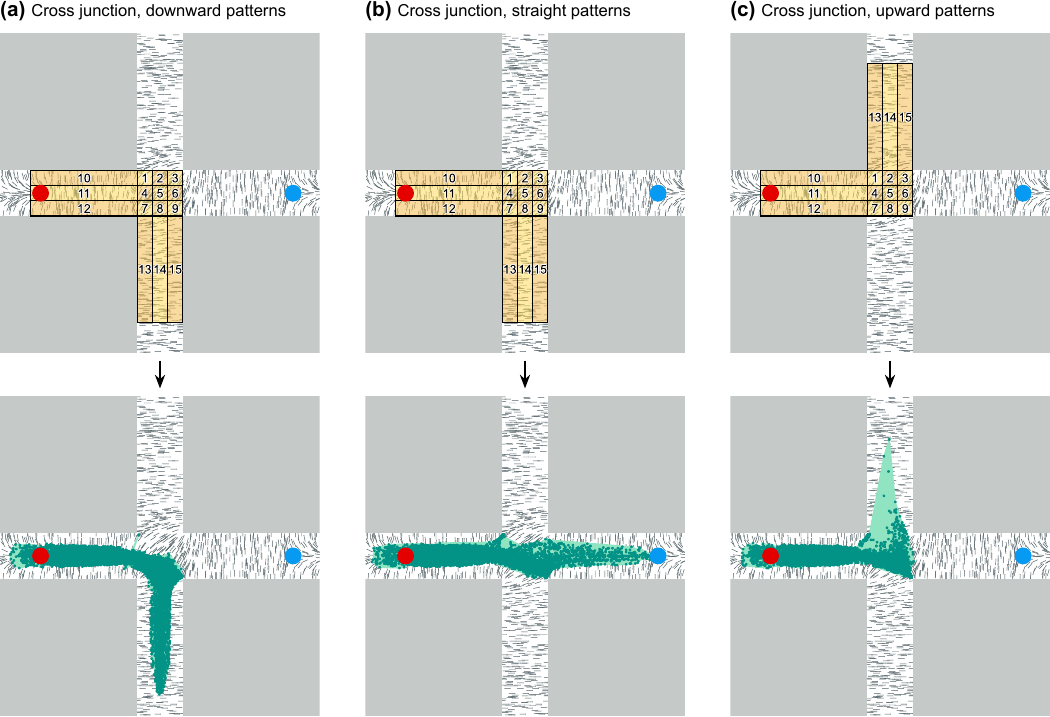}
	\caption{Reachability analysis with global control pattern sets on the cross junction. The top row displays the initial configurations and all available pattern primitives, and the bottom row shows the corresponding 20-step reachable set estimates. The yellow pattern primitives are long rectangles and smaller squares; each numbered primitive may be independently enabled or disabled at each control step. For (a) downward and (b) straight control, the estimated reachable sets extend from the initial left channel to the far ends of the bottom and right channels. The reached positions are dense, indicating that these global patterns are suitable for control. (c) For upward control, the estimated reachable set also includes the top channel, but the point density significantly drops there. Thus, top-channel goals are reachable only for a small fraction of activity sequences, making control difficult but not impossible.}
	\label{fig:reach_sets_global}
	\end{figure*}

	We now use reachable sets to compare how different control pattern sets change defect mobility. The resulting maps show that activity patterns do not merely push defects along prescribed directions; they interact with the director field and walls to create anisotropic, geometry-dependent control landscapes. Figs.~\ref{fig:reach_sets_local} and \ref{fig:reach_sets_global} show the control pattern sets under consideration, along with their resulting reachable sets.

	We consider two classes of pattern sets. Local sets contain small rectangular activity strips defined relative to the current defect position, while global sets contain activity patterns at fixed positions relative to the simulation lattice. Global sets include both small junction-scale and long channel-scale pattern primitives. In all pattern figures, yellow marks regions where activity is switched on; outside those regions, the fluid evolves with no applied active stress.

	For local sets, exactly one pattern can be enabled at each control step, whereas for global sets multiple primitives can be applied at the same time. In reinforcement-learning terminology, local sets define discrete action spaces, while global sets define multibinary action spaces. The resulting reachable sets differ strongly, with direct implications for action-space design.

	\subsection{Local Control Pattern Sets}
	\label{subsec:reach_local}

	We begin with the free geometry, where the initial director field is vertical across the whole domain. This anisotropy breaks rotational symmetry, both in the system and in the reachable set. In the top panel of Fig.~\ref{fig:reach_sets_local}(a), each arm of the yellow cross is a distinct 4-way local pattern, and the bottom panel shows the estimated reachable set. The up and down local patterns have little effect on defect motion, so the reachable set has only a narrow vertical extent. By contrast, the left and right patterns couple more effectively to the initial director field and spread the reachable set horizontally.

	The local pattern sets are deliberately restrictive. At each control step, the active region is placed relative to the current $+1/2$ defect position rather than fixed in the laboratory frame. These controls therefore test whether defect motion can be generated by local actuation alone, without pre-patterning the downstream channel or modifying the surrounding geometry as a whole.

	The anisotropy in Fig.~\ref{fig:reach_sets_local}(a) is not simply a geometric property of the pattern set. The same rectangular activity patterns would produce a different response in a different background director field. Here the vertical far-field director makes horizontal actuation more effective because it more readily generates director distortions and active flows that translate the comet-shaped defect. Vertical strips, however, mostly reinforce an unfavorable local alignment and produce little net displacement over a control interval.

	The 8-way local set in Fig.~\ref{fig:reach_sets_local}(b) gives a different result. It contains the four patterns above, plus four diagonal patterns, and the reachable set now extends much farther in the vertical direction while its horizontal extent changes little. Diagonal patterns are therefore crucial for full defect mobility in the free geometry. They can locally tilt the director field upward or downward and permit defect flow, whereas vertically oriented patterns fail to do this effectively. Intuitively, when the angle between the local director and the long axis of the rectangular pattern is too large, active stress cannot strongly reorient the director, inhibiting defect motion.

	The diagonal activity patterns effectively introduce oblique shear and bend distortions around the defect core. These distortions rotate the defect polarization and let subsequent activity steps convert local forcing into vertical displacement. Thus, the improvement from four to eight patterns is not only a larger action count; it also changes which director deformations are accessible. The same pattern-dependent response appears later in the trained-controller trajectories of Fig.~\ref{fig:perf_free_local8}.

	In the cross junction of Fig.~\ref{fig:reach_sets_local}(c), the same local strategy fails. The spatially varying initial director field and homeotropic wall anchoring strongly constrain mobility. Since the 4-way local set is already limited, we apply only the 8-way set to this geometry.

	The cross junction is a stronger test because local actuation must compete with two sources of constraint. First, the homeotropic walls impose a preferred director orientation near each channel boundary. Second, the diagonal director field in the intersection biases the defect toward some exits and away from others. A local activity pattern centered on the defect can perturb the director near the core, but it cannot reorganize the director field across the full junction.

	Even this expanded set gives very limited mobility: the reachable set remains confined to the left channel, while the other three channels and even the junction itself are unreachable within the 20-step horizon. The diagonally oriented director field in the junction, together with wall anchoring, acts as a reachability barrier. Local control pattern sets are therefore insufficient for steering the defect through this microchannel junction.

	This failure motivates the global pattern sets considered next. If the junction-scale director field is itself part of the barrier, then successful control requires activity patterns that act on the channel and intersection, not only on the immediate defect neighborhood.

	\subsection{Global Control Pattern Sets}
	\label{subsec:reach_global}

	Global control patterns produce much larger reachable sets. We first consider the downward global set in the cross junction (Fig.~\ref{fig:reach_sets_global}(a)), where activity pattern primitives are placed across the left and bottom channels. Each yellow quadrilateral outlined in black may be enabled or disabled independently, giving $2^{15}$, or more than $3 \times 10^{4}$, possible patterns.

	The reachable set now covers the intersection and extends through the bottom channel. Global control pattern sets therefore greatly expand reachable regions in confined geometries, making the defect mobile across both the left and bottom channels. The sample points are also relatively uniform across the reachable set, which helps RL training because irregular patchworks of defect positions produce more complicated relationships between actions and rewards.

	The straight global set gives similar results (Fig.~\ref{fig:reach_sets_global}(b)). The reachable set includes the intersection and the right channel, although positions become somewhat less dense in the right half. The upward global set gives a notably different result (Fig.~\ref{fig:reach_sets_global}(c)). This asymmetry in mobility is expected because the initial director field in the junction is not up-down symmetric. Although many reached positions appear in the junction, the top channel is sparsely populated. In fact, a detailed check of the $10^4$ random control steps reveals that the original $+1/2$ defect never enters the top channel.

	The top-channel points therefore have a different origin. Some activity patterns create a new defect pair at the northeast corner of the junction, and the sparse top-channel points are reached by the newly created $+1/2$ defect. As explained in Appendix~\ref{sec:app_algos}, we retain these positions in the estimated reachable set.

	This distinction matters because the reachable set of the original defect can be smaller than the effective reachable set of any controllable $+1/2$ defect. Pair creation can enlarge the accessible region by changing defect identity. New defect pairs also appear for some straight-global sequences; in Fig.~\ref{fig:reach_sets_global}(b), the original defect can often cross the junction intact, but defect creation is still common. For control tasks with goals in the right channel, suppressing unwanted defect creation may be important.

	\subsection{Consequences for Defect Control}
	\label{subsec:reach_implications}

	These reachability analyses reveal how geometry, director structure, and the control pattern set shape defect mobility. One immediate result is that the pattern set strongly controls the size and shape of the reachable set. In the free geometry, adding diagonal local patterns greatly expands the reachable region. In the cross junction, by contrast, large regions remain inaccessible unless global patterns are used. Thus, local patterns suffice in open domains, but global patterns are needed for transport through intersections.

	Even global patterns can fail for some director configurations. It may be impossible to guide one chosen defect through every region of an arbitrary microchannel network, a conclusion that is consistent with the strong anisotropy of nematic materials. Pair creation can change that conclusion: a newly created $+1/2$ defect may reach positions that the original defect cannot. This suggests a control strategy in which a new defect is deliberately created when the target lies in an otherwise inaccessible region. In that sense, a control task may remain solvable even when the original defect cannot pass through a junction.

	These dynamics give RL controllers room to learn counterintuitive strategies. Reachability analysis therefore gives a clear physical picture of the control landscape before any learning is performed. It identifies which failures are due to poor controller design and which are due to genuine mobility constraints imposed by the active nematic field configuration. We introduce this method of analysis here not only for active nematics, but also for other soft and active matter systems with complex dynamics.

	Before proceeding, we note that the activity intensity ($\alpha_0$ in Table~\ref{tab:lbm-params} of Appendix~\ref{sec:app_be_eq}) is kept low to prevent continual chaotic creation and annihilation of defect pairs. If activity is too large, the applied patterns produce high defect densities and active turbulence rather than controlled motion. One could define a different task in which the only goal is to create some $+1/2$ defect near the target; a controller might then inject strong activity near the goal until a defect pair appears, and then attempt to stabilize the new $+1/2$ defect. Such a task could be useful for studying defect statistics or turbulence, but this is not the main aim of the present study.

	Our aim is controlled manipulation of existing defects. This is more relevant for microfluidic devices, where one would ideally apply low active stress, guide the desired defect, and avoid creating unwanted defects elsewhere. Still, occasional pair creation remains possible at our activity level, with important consequences for control, as shown in the next section.

	\section{Controller Performance Results}
	\label{sec:controller_perf}

	\begin{figure*}[t]
	\centering
	\includegraphics[width=\textwidth]{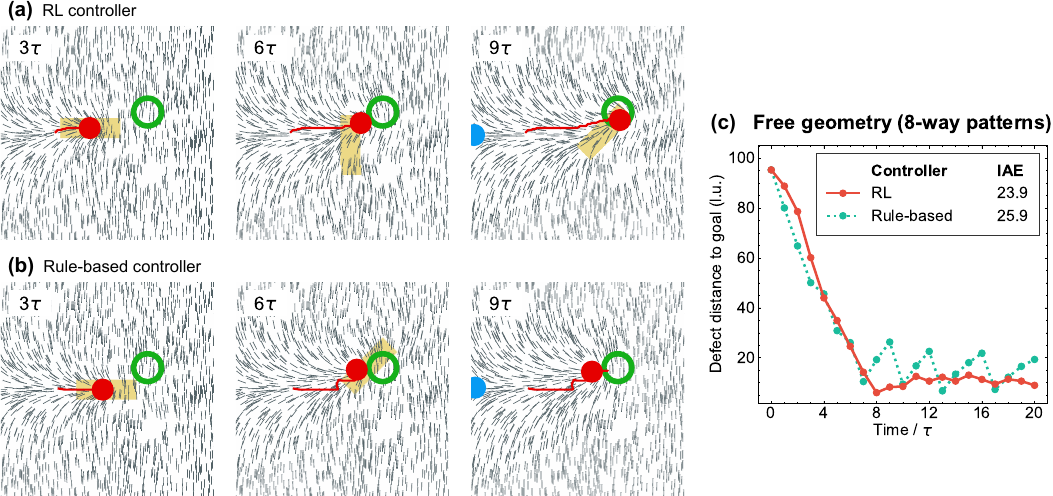}
	\caption{Defect trajectories in the free geometry using the 8-way local pattern set, generated by (a) the RL controller and (b) the rule-based controller. Both controllers guide the $+1/2$ defect to the goal (green ring), but the RL controller better stabilizes it there after arrival. Meanwhile, the rule-based controller overshoots and produces large oscillations around the goal, unable to effectively counteract the attraction due to the approaching $-1/2$ defect (blue). Snapshots are zoomed in for clarity. The error plot in (c) compares defect-to-goal distances vs. time under the two controllers, where the different stabilization behaviors can be clearly seen. Lower IAE scores are better, and distances are given in lattice units.}
	\label{fig:perf_free_local8}
	\end{figure*}

	The reachable sets identify where control should be possible. The next question is whether a closed-loop controller can actually find fast, stabilizing activity sequences in these high-dimensional action spaces. For example, a 20-step episode with a global pattern set has $(2^{15})^{20} \approx 2 \times 10^{90}$ possible control sequences. Despite this dimensionality, the trained RL controllers solve their training tasks, outperform the baselines, and generalize well to a larger test geometry. Their trajectories also show how defect creation, defect annihilation, and persistent free energy distortions shape active nematic control.

	To quantify each trajectory, we use the time-averaged \emph{integral of absolute error} (IAE), a standard control metric \cite{Graham_Lathrop_1953, Kirk_2012}. During a test episode, we apply a trained controller to the hybrid LBM simulation and record the $+1/2$ defect position over time. The time-averaged IAE is the defect-to-goal distance averaged over the episode:
	\begin{equation}
		\frac{1}{N \tau} \int_{0}^{N \tau} \| \textbf{x}_{d} (t) - \textbf{x}_{g} \| \, dt,
	\end{equation}
	where $\textbf{x}_{d} (t)$ is the (2D) defect position in lattice units at time $t$, $\textbf{x}_{g}$ is the goal position, and $N \tau$ is the episode duration ($N$ is the number of control steps per episode, while $\tau$ is the aforementioned control step duration). Lower IAE values correspond to better performance. In practice, we approximate the integral by a discrete sum over the defect positions observed at the end of each control step $\{\tau, 2 \tau, \dots, N \tau \}$ via
	\begin{equation}
		\frac{1}{N} \sum_{k = 1}^{N} \| \textbf{x}_{d} (k \tau) - \textbf{x}_{g} \|.
	\end{equation}

\begin{figure*}[t]
	\centering
	\includegraphics[width=\textwidth]{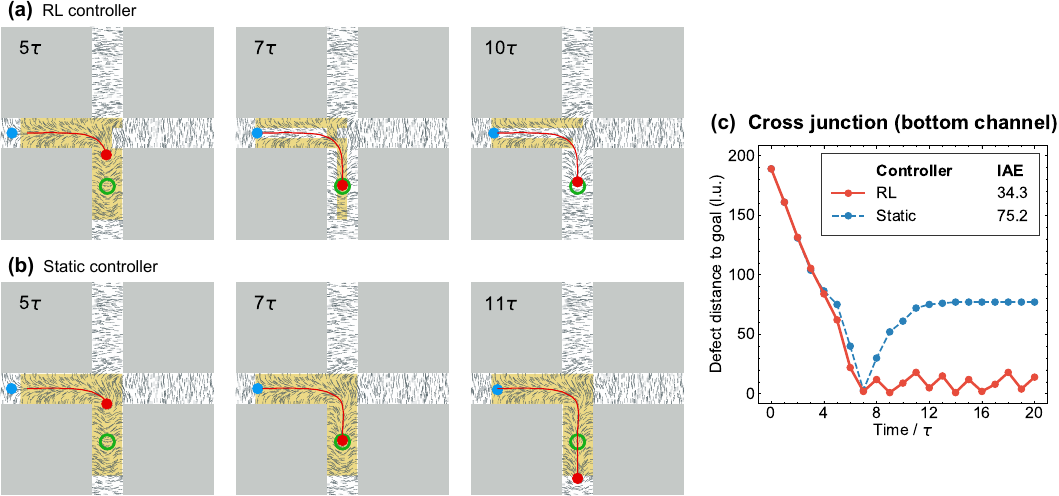}
	\caption{Defect trajectories on the cross junction with a bottom-channel goal, generated by (a) the RL controller and (b) the static controller. Under static control, the defect passes through the goal but continues to the end of the bottom channel. Under RL control, the defect reaches the goal at the same speed, but remains near it for the rest of the episode thanks to the RL controller's dynamic stabilization. The error plot and IAE scores in (c) quantify the superior performance of RL control over static control.}
	\label{fig:perf_cross_down}
	\end{figure*}

\begin{figure*}[t]
	\centering
	\includegraphics[width=\textwidth]{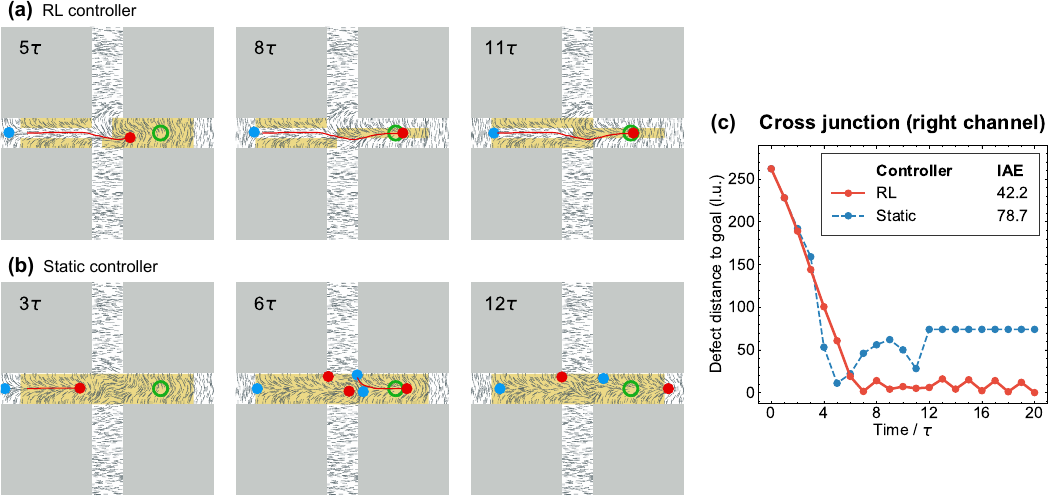}
	\caption{Defect trajectories on the cross junction with a right-channel goal, generated by (a) the RL controller and (b) the static controller. Under static control, two new defect pairs form in the junction, producing chaotic motion that eventually annihilates the original $+1/2$ defect. For the RL controller, defect creation was heavily penalized during training, and the learned policy steers the original defect intact to the goal without creating unwanted defects. The error plot and IAE scores in (c) show that the RL controller solves the task and achieves much lower error than the static controller.}
	\label{fig:perf_cross_straight}
	\end{figure*}

\begin{figure*}[t]
	\centering
	\includegraphics[width=\textwidth]{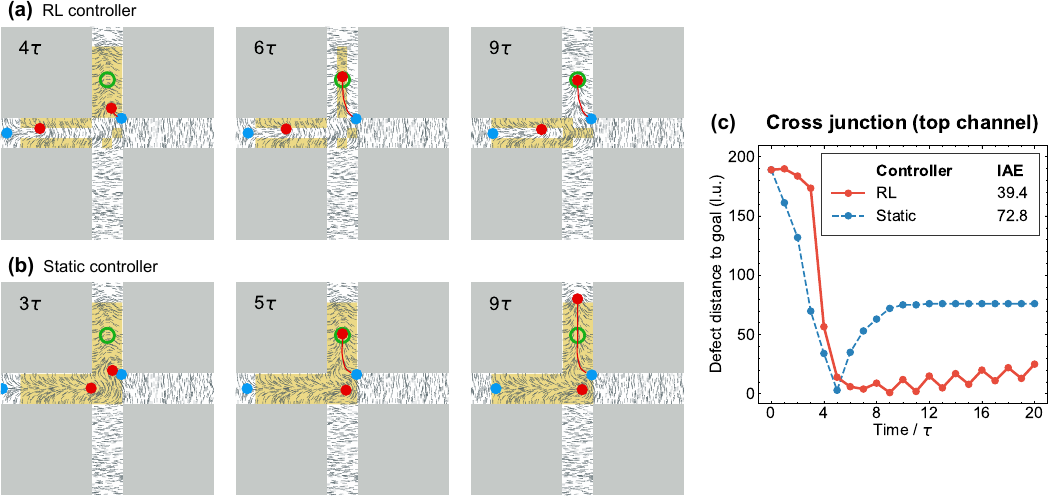}
	\caption{Defect trajectories on the cross junction with a top-channel goal, generated by (a) the RL controller and (b) the static controller. With static control, a defect pair forms at the northeast corner of the intersection, and the new $+1/2$ defect travels through and beyond the goal. For the RL controller, defect creation was \emph{not} penalized during training; the learned policy creates a pair at the same location, switches to the new $+1/2$ defect, and stabilizes it at the goal. If defect creation is forbidden during training, the RL controller fails to find a control sequence that pushes the original $+1/2$ defect up through the junction, suggesting that intact passage is impossible for this director field and activity level. The error plot and IAE scores in (c) show again that RL solves the control problem.}
	\label{fig:perf_cross_up}
	\end{figure*}

	On a plot of defect-to-goal distance versus time, the time-averaged IAE is simply the mean value of the curve, making it a natural metric for the minimum-time control problems studied here. The RL training reward has a different form: reward shaping improves convergence but produces a complex function that is harder to interpret (see Appendix~\ref{sec:app_rl_params} for details). IAE is therefore better suited for evaluating trained controllers. In Figs.~\ref{fig:perf_free_local8}--\ref{fig:perf_cross_up}, we show selected trajectories and report error plots and IAE scores. Additional T-junction trajectories appear in Appendix~\ref{sec:app_additional_results} (Figs.~\ref{fig:perf_t_down} and \ref{fig:perf_t_up}), and movies are provided in the Supplementary Material.

	\subsection{Free Geometry}
	\label{subsec:control_perf_free}

	Despite its simplicity, the free geometry is a good proving ground. Figs.~\ref{fig:perf_free_local8}(a) and \ref{fig:perf_free_local8}(b) show trajectories from the trained RL controller and the rule-based controller, both using the 8-way local pattern set. The goal (green ring) lies slightly above the domain centerline.

	Because the target position lies well inside the reachable set, this case is not mainly a test of access. Instead, it isolates a different part of the control problem: once a defect reaches the goal, the controller must keep it there despite continuing active motion, defect-defect interactions, and the anisotropic response of the director field.

	Both controllers bring the defect to the goal, but their stabilization behavior differs. Fig.~\ref{fig:perf_free_local8}(c) displays the defect-to-goal distance over time, and the IAE scores show that RL control performs better. The rule-based controller prevents the defect from straying too far, but it produces large oscillations in the second half of the episode; the RL controller reduces those oscillations by roughly an order of magnitude.

	This difference is important because the IAE penalizes both slow arrival and poor stabilization after arrival. A controller that reaches the target quickly but then overshoots repeatedly can still have a large IAE. The free-geometry task therefore separates two aspects of minimum-time position control: rapid approach to the goal and suppression of residual motion once the goal has been reached.

	The control sequences explain this difference. The rule-based controller chooses the empty pattern when the defect is close to the goal. Even with no applied activity, however, the $+1/2$ defect moves due to attraction to the ${-1/2}$ defect (blue), which enters from the left through the periodic boundary. Some activity is therefore needed to counteract this attraction, but the left and right patterns inject too much activity and cause the $+1/2$ defect to flow rapidly in the corresponding direction.

	The rule-based controller fails because its implicit model of the defect dynamics is too local. It assumes that the selected strip determines the next displacement, but in the simulation the defect also responds to the surrounding director field and to the other defect in the periodic domain. Near the goal, these additional forces become comparable to the displacement produced by one control step, so a greedy geometric rule fails to provide stable feedback control.

	The vertical and diagonal patterns behave differently: they produce active stresses that counteract the attraction without driving strong unwanted motion. The RL controller uses this incompatibility between selected patterns and the anisotropic director field to prevent runaway motion. By contrast, the rule-based controller uses only the left, right, and empty patterns after the defect first reaches the goal. It approaches the target quickly at first, but then repeatedly overshoots, while the RL controller uses a broader subset of patterns and keeps the defect close to the goal.

	The physical takeaway is that activity patterns can damp defect motion as well as enhance it. Even in the simple free geometry, successful control is not equivalent to always choosing the pattern that points most directly toward the target. The controller must also learn which patterns weakly actuate, strongly actuate, or effectively brake the defect within the broader director environment. This distinction becomes more important in junctions, where walls and broken-symmetry director fields further restrict the available motions.

	\subsection{Cross Junction}
	\label{subsec:control_perf_cross}

	Controller results for the cross junction show how geometry and anchoring complicate steering. The rule-based controller uses local patterns, which produce very small reachable sets in junction geometries, so we instead use static controllers with fixed global patterns as more appropriate and challenging benchmarks.

	We begin with the goal in the bottom channel (Fig.~\ref{fig:perf_cross_down}). Under static control, constant application of the full downward global pattern steers the defect through the goal, after which it continues to the lower outlet. Under RL control, the defect initially follows a similar path and travels downward through the junction at essentially the same speed. The difference in behavior appears near the goal, where the RL controller repeatedly applies and removes a long rectangular activity strip along the center of the bottom channel to dynamically stabilize the defect. The RL controller therefore solves the bottom-channel task and achieves a much lower IAE score than the static controller (Fig.~\ref{fig:perf_cross_down}(c)).

	Before examining control tasks in the right and top channels, we note a design choice in the global control pattern sets. The full downward pattern is divided into small squares inside the junction and long strips inside the channels. Though one may employ equal-sized squares everywhere, long channel strips reduce the action-space dimension without shrinking the reachable set. Long patterns thus regularize the control problem.

	Breaking the strips into smaller squares would allow finer control, but the free-geometry results already show that such localized RL control is possible. In the cross junction, we use long strips to test a different action-space topology. This actually makes stabilization harder, since no single action can simply hold the defect at the goal. Some oscillatory motion becomes inevitable. Even so, the RL controllers converge to successful policies.

	Fig.~\ref{fig:perf_cross_straight} shows the right-channel task. At first glance this should be easier than the bottom-channel task, since the defect only needs to move straight through the junction. However, the global pattern set makes the problem more subtle. Under static control, multiple defect pairs are created and annihilated in the junction: constant application of the full pattern produces enough active stress to create two new pairs, and the original $+1/2$ defect then annihilates with a new ${-1/2}$ defect near the intersection.

	This behavior highlights the danger of static patterns. Many straight-global patterns can trigger unwanted defect creation and annihilation, so the RL controller cannot simply activate most primitives to maximize speed. It must guide the defect through the junction intact while still minimizing travel time. To encourage this behavior, we penalize new defect creation during training. The resulting policy avoids applying too much activity to the problematic northeast corner of the junction, and the defect follows a slightly curved path into the right channel without annihilating (Fig.~\ref{fig:perf_cross_straight}(a)).

	In Fig.~\ref{fig:perf_cross_straight}(c), the static-controller error is computed using the $+1/2$ defect nearest the goal, since the original $+1/2$ defect is annihilated early. This artificially improves the static-controller IAE; even under this generous comparison, RL control performs much better.

	Top-channel goals produce other interesting dynamics (Fig.~\ref{fig:perf_cross_up}). Under static control with the full upward pattern, the original $+1/2$ defect rapidly approaches the junction, but it does not enter the top channel. Instead, a defect pair forms at the northeast corner, and the new $+1/2$ defect flows upward past the goal. This agrees with the reachability analysis: we found no sampled sequence in which the original defect entered the top channel. When defect creation is forbidden during RL training, the controller also fails to steer the original defect up through the junction, even after several days of training.

	Pair creation provides an alternative control strategy. We trained another RL controller to manipulate whichever $+1/2$ defect is nearest the goal, without penalizing defect creation. This lets the controller switch focus from the original defect to a newly created one. In Fig.~\ref{fig:perf_cross_up}(a), the controller does exactly this: it relinquishes control over the original defect, switching instead to the new $+1/2$ defect and stabilizing it at the top-channel goal, earning a low IAE score (Fig.~\ref{fig:perf_cross_up}(c)).

	This demonstrates the importance of reward design. Steering the original defect into the top channel is extremely difficult or even impossible for this director field and activity level. By exploiting pair creation, the RL controller can still solve the task with a different $+1/2$ defect. Allowing or forbidding pair creation during training thus leads to novel, counterintuitive control strategies.

	One could try to handcraft a global-pattern rule-based controller. For example, one might apply the full global pattern until the defect reaches the goal, then switch to a local stabilizing rule. Such controllers may work for one cross junction, but it is unclear how to adapt them to arbitrary channel geometries. In contrast, deep RL can autonomously discover control strategies that are effective across different geometries and control tasks.

	\subsection{Cross Maze Test Geometry}
	\label{subsec:control_perf_cross_maze}

	\begin{figure*}[t]
	\centering
	\includegraphics[width=\textwidth]{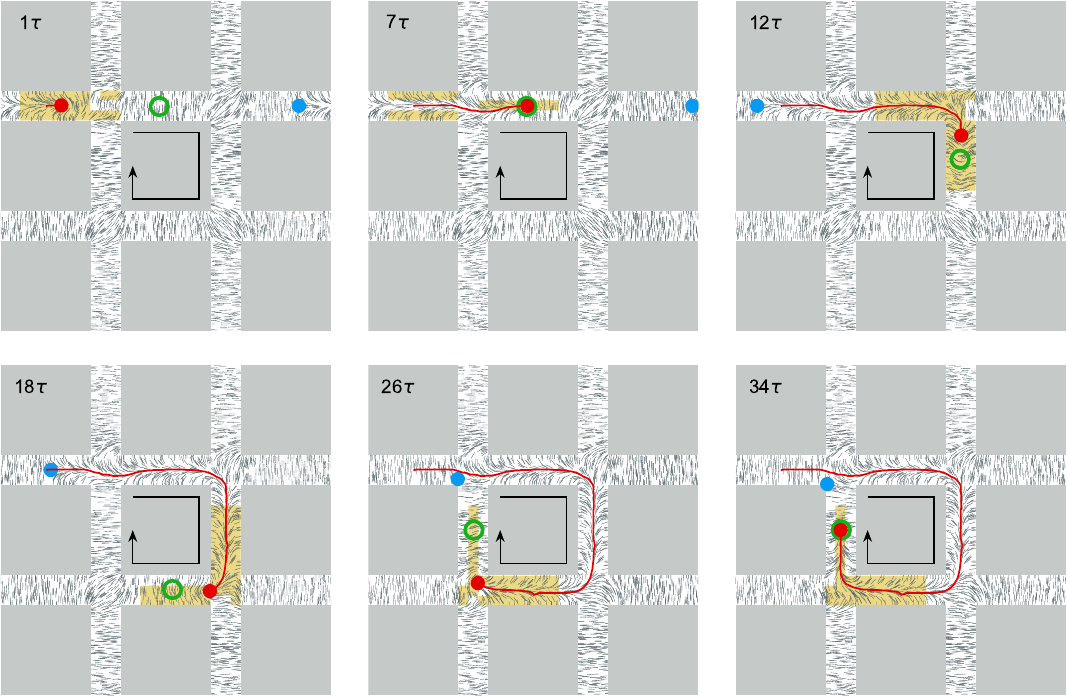}
	\caption{Defect trajectory generated by the meta-controller on the cross maze test geometry. The defect follows a clockwise circuit, indicated by the looping arrow, while the changing goal position serves as a sequence of waypoints. The meta-controller combines RL controllers trained only on the simpler cross junction geometry; their weights are frozen, and no training or fine-tuning is performed on the maze. The meta-controller succeeds in guiding the defect around the full circuit.}
	\label{fig:rl_cross_maze_cw}
	\end{figure*}

	The strong performance on the training geometries suggests that the RL policy networks have enough capacity for these control problems. The next question is whether trained controllers can be reused in unseen geometries, thereby testing the generality and adaptability of the learned policies.

	For this purpose, we use the cross maze test geometry (Fig.~\ref{fig:rl_cross_maze_cw}). It consists of four overlapping cross junctions, each the same size as the training geometry, leading to a full lattice size of 660-by-660 sites. The director orientation in each intersection is chosen so that the $+1/2$ defect can navigate clockwise without creating new defect pairs. To steer the defect around a clockwise loop, we use a sequence of waypoints: once the defect reaches one waypoint, the goal switches to the next, producing a multi-step manipulation task.

	The trajectory in Fig.~\ref{fig:rl_cross_maze_cw} is generated by a meta-controller that alternates among three trained cross-junction RL controllers, corresponding to goals in the bottom, right, and top channels. At each step, it selects the controller based on the current maze quadrant and the relative positions of the defect and goal. Note that the RL weights are frozen after cross-junction training; no further training or fine-tuning is done on the maze.

	The meta-controller uses a simple rotation algorithm. Because the trained controllers assume that the defect starts in the left channel, the relevant quadrant of the maze is first extracted as a 420-by-420 grid, matching the cross junction training geometry. This window is rotated so that the defect appears to enter from the left channel, and the appropriate trained RL controller is then applied in this rotated frame. After the controller generates an activity pattern, the pattern is rotated back and applied to the maze. Once the defect reaches the current waypoint, the meta-controller moves to the next quadrant and repeats the procedure.

	As an example, consider Fig.~\ref{fig:rl_cross_maze_cw} near $t = 18 \tau$, when the defect passes through the southeast junction. The algorithm extracts a 420-by-420 window aligned with the southeast corner. Since the defect enters from the top channel, the window is rotated 90 degrees counterclockwise. In the rotated frame, the defect enters from the left and the goal lies in the bottom channel, so the downward controller is selected. The generated activity pattern is then rotated 90 degrees clockwise and applied to the southeast quadrant.

	The director fields in the maze intersections were chosen so that each quadrant can be aligned to the training junction by rigid rotation alone, without reflections. For arbitrary initial director configurations, the alignment algorithm could be extended to include reflections as well. This rotation procedure allows controllers trained on one entrance channel to be used for all four directions, reducing required training times and providing a simple way to reuse smaller controllers in larger geometries.

	The resulting clockwise trajectory is shown in Fig.~\ref{fig:rl_cross_maze_cw}. The ${-1/2}$ defect positions, and therefore the initial director fields, differ between the training junction and the maze quadrants. Despite this mismatch, the meta-controller steers the $+1/2$ defect through each waypoint and around the maze to the final goal, demonstrating remarkable performance.

	However, the meta-controller underperforms on some maze tasks. For example, it fails to complete a counterclockwise loop (see the Supplementary Material). In this direction, the upward global controller must be used at most junctions to create new defect pairs, requiring repeated changes of defect focus and making the task harder. Though the meta-controller succeeds at the first two intersections along the counterclockwise loop, it struggles at the third.

	These tests show both the generality of the trained controllers and their limits. Separate cross-junction policies can be composed into a higher-level controller, allowing defects to be steered clockwise around the maze in a multi-step maneuver. On the other hand, the counterclockwise failure shows that the effectiveness of trained policies may depend on compatible director configurations and junction orientations.

	\subsection{Evolution of Free Energy Fields}
	\label{subsec:control_free_energy_analysis}

	\begin{figure*}[t]
	\centering
	\includegraphics[width=\textwidth]{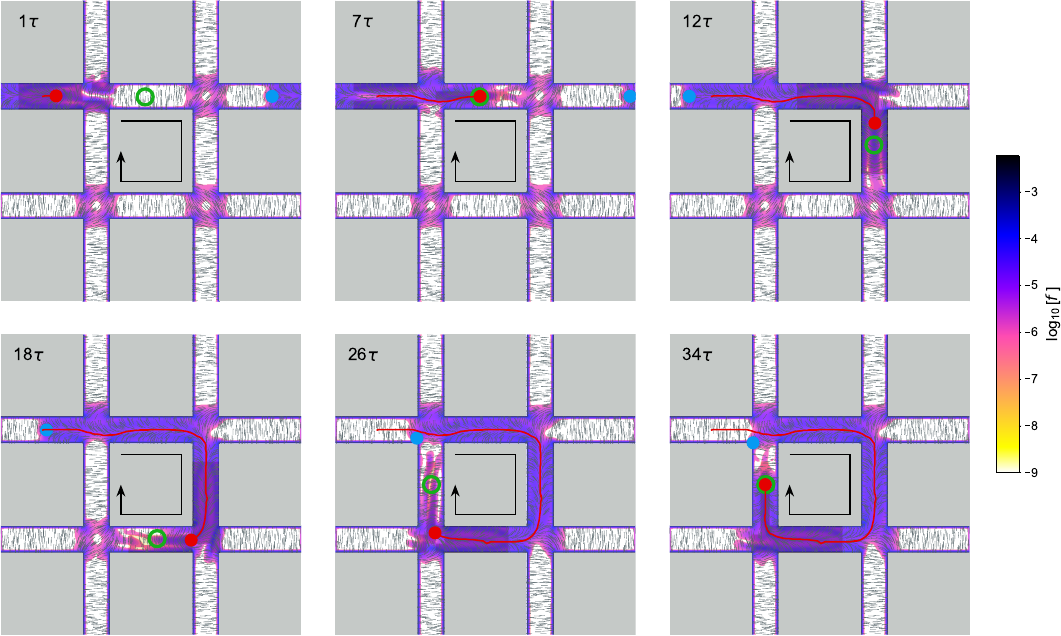}
	\caption{Free energy field (in lattice units) for the cross maze during the same clockwise trajectory as in Fig.~\ref{fig:rl_cross_maze_cw}. Energies are plotted on a logarithmic scale. Activity patterns immediately raise the local free energy, but these high-energy director configurations persist even after activity is switched off. Large regions of the channel interiors and junctions remain altered in the wake of the traveling $+1/2$ defect. Meanwhile, the $-1/2$ defect lowers the surrounding free energy as it drifts, partially erasing the path history of the controlled defect. Such memory effects influence the mobility of later defects and could be used for microfluidic information storage and processing.}
	\label{fig:fe_maze_cw}
	\end{figure*}

	We gain further insight by tracking the Landau-de Gennes free energy during the maze trajectory. Fig.~\ref{fig:fe_maze_cw} shows the same clockwise trajectory as Fig.~\ref{fig:rl_cross_maze_cw}, now with the free energy field plotted on a logarithmic scale. Free energy is largest near the channel walls, as expected from infinite homeotropic anchoring, while the channel bulk shows clear history dependence. Lines of abrupt change in free energy coincide with $+1/2$ and ${-1/2}$ defect positions, forming sharp gradient walls that move downstream with their corresponding defects.

	As the $+1/2$ defect moves from one junction to the next, its path is etched into the director field. The resulting distortions produce high-energy regions, so the system displays memory-like behavior: channels and junctions remain in elevated energy configurations long after the guided defect has passed. The ${-1/2}$ defect can partially erase this memory. When it enters these high-energy regions from $t = 12 \tau$ onward, it lowers the surrounding free energy as it moves. Traveling defects therefore reshape the free energy landscapes of channels and intersections.

	The persistence of these energy changes shows that the controller is not merely translating a defect through a fixed background. Each applied activity pattern locally modifies the balance between elastic relaxation, wall anchoring, and active forcing. After the activity is removed, the director field does not immediately return to its initial low-energy configuration, because the defect has transported distortions across the channel and left them constrained by the surrounding walls. The next control action therefore acts on a modified material state. This history dependence is one reason why the same activity pattern can produce different defect motion at different stages of a trajectory.

	These visualizations show that activity patterns do more than move defects. They alter the material state through which later defects travel. This extends the finding of Zhang et al. \cite{Zhang_Mozaffari_de_Pablo_2022}: once a $+1/2$ defect passes through a junction, the altered director field can block a second $+1/2$ defect. Such history-dependent dynamics are relevant for microfluidic logic, where they may enable fluidic computing primitives analogous to transistor gates. Because ${-1/2}$ defects can erase parts of the path history, deliberate control of defects of both signs could tailor energy landscapes, paving the way for active nematic information storage and processing.

	\subsection{Optimality Analysis}
	\label{subsec:control_optim_analysis}

	\begin{figure*}
	\centering
	\includegraphics[width=\textwidth]{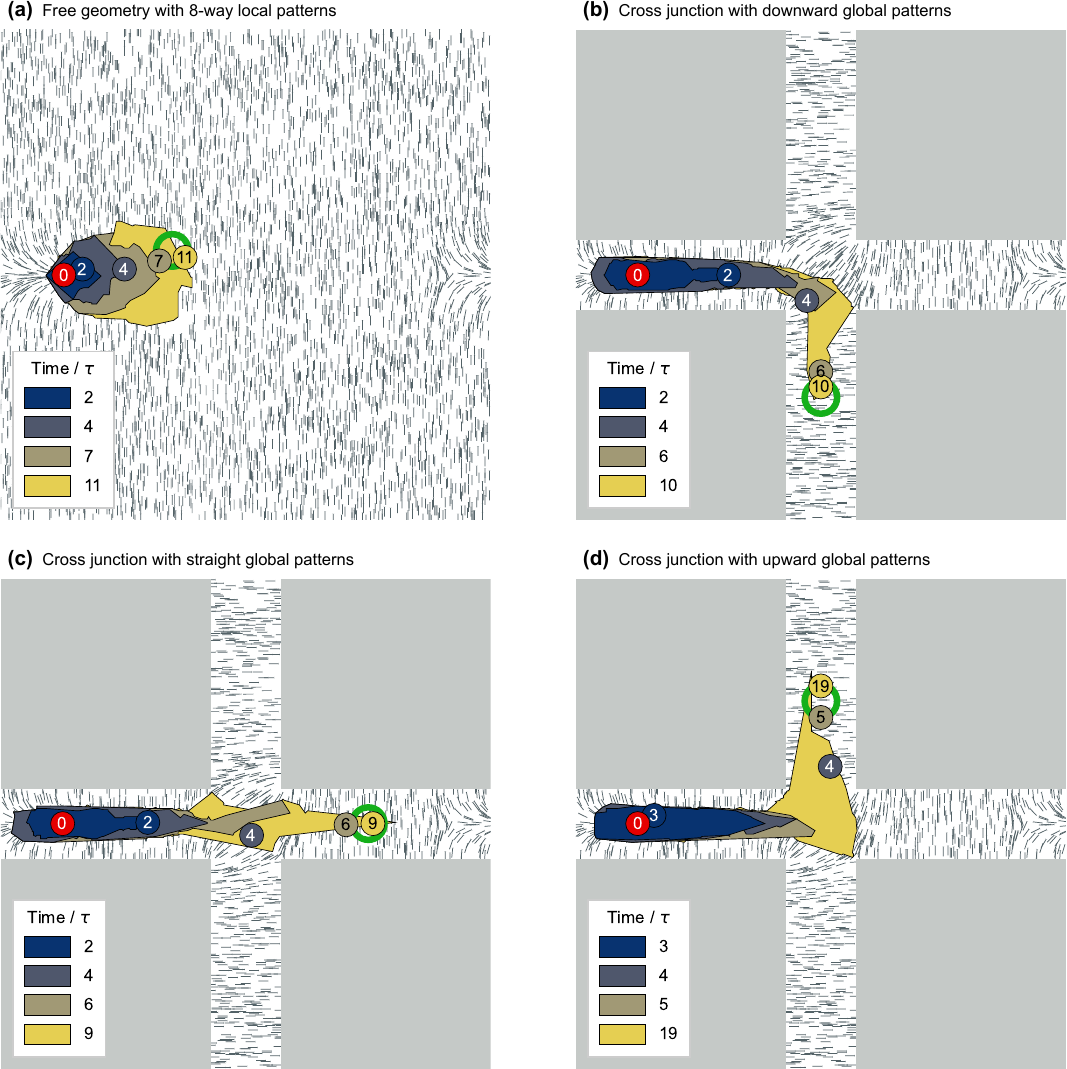}
	\caption{Optimality visualizations for RL-generated trajectories in (a) the free geometry with the 8-way local pattern set, and in the cross junction with goals in the (b) bottom, (c) right, and (d) top channels. Time-stepped reachable sets for different episode lengths are shown in multiples of the control step duration $\tau$. Defect positions from RL trajectories are overlaid in matching colors, with control step numbers inscribed. The defect positions consistently reach or exceed the estimated reachable set boundaries at the corresponding times, indicating near-optimal behavior for these minimum-time control problems.}
	\label{fig:optimality_analysis}
	\end{figure*}

	Finally, we ask whether the learned trajectories are merely local optima or whether they are close to the fastest trajectories allowed by the reachable landscape. Global optimality is difficult to verify exactly, but reachable sets provide a diagnostic. We therefore use time-stepped reachable sets to compare learned control policies with the fastest allowable trajectories.

	The \emph{reachable set in $k$ control steps}, or \emph{$k$-reachable set}, is the collection of all defect positions reachable from the initial position within $k$ control steps for a given control pattern set \cite{Sontag_1998}. As $k$ increases, these sets grow, with the $k$-reachable set contained in the $(k+1)$-reachable set.

	For a minimum-time control problem, an optimal trajectory should lie on the boundary of the $k$-reachable set at step $k$, at least until the goal becomes reachable. This follows from the definition of reachability: if the defect lies inside the boundary at step $k$, then another trajectory must have moved farther in the same time, making the original trajectory suboptimal. This is one manifestation of the Pontryagin maximum principle \cite{Kirk_2012, Sontag_1998, Pontryagin_1962}. By overlaying RL trajectories on time-stepped reachable sets, we can therefore judge minimum-time optimality by seeing how closely the defect approaches the expanding boundary.

	Fig.~\ref{fig:optimality_analysis} shows this procedure for the free-geometry controller and the three cross-junction controllers. The trajectories are the same as those in Figs.~\ref{fig:perf_free_local8}--\ref{fig:perf_cross_up}. Controlled $+1/2$ defect positions are shown as numbered circles, and for each representative step, the $k$-reachable set and the RL defect position at step $k$ use the same color. The defect positions and reachable sets come from different procedures: the defect positions are generated by trained RL controllers during one episode, while the $k$-reachable sets are estimated by random sampling and concave-hull generation with maximum episode length $k$, as described in Section \ref{sec:method} and Fig.~\ref{fig:reach_set_est_method}. At $t = 0$, the reachable set is the initial position only; at $t = 20 \tau$, the sets recover the full reachable sets in Figs.~\ref{fig:reach_sets_local} and \ref{fig:reach_sets_global}.

	In the free geometry (Fig.~\ref{fig:optimality_analysis}(a)), the RL trajectory lies along the growing reachable-set boundaries at all shown steps. This strongly suggests that the pattern sequence is near optimal. On the other hand, the bottom-channel cross-junction task requires more care to interpret (Fig.~\ref{fig:optimality_analysis}(b)). At $t = 2 \tau$, the defect again lies on the estimated boundary. At $4 \tau$ and $6 \tau$, however, it appears just outside the boundary.

	This may seem to contradict the definition of a reachable set, since a defect should not be able to lie outside the set of positions reachable in a given number of control steps. However, recall that our concave-hull estimates are subsets, and therefore underestimates, of the true $k$-reachable sets. The RL controller has driven the defect beyond these estimated hulls, but the defect must still lie within the true reachable sets. This is expected because reachability analysis samples $10^4$ random control steps, while RL training runs for $10^5$ steps or more, allowing the controller to explore more of the true reachable set. As a result, defects beyond the estimated boundaries are evidence of strong performance.

	The right-channel task shows the same phenomenon (Fig.~\ref{fig:optimality_analysis}(c)). The defect travels beyond the estimated boundaries at steps 4 and 6. In the top-channel task (Fig.~\ref{fig:optimality_analysis}(d)), the original defect stays near its initial position at early times. At $t = 4 \tau$, a new defect pair appears well outside the estimated boundary, and the RL controller quickly guides the new $+1/2$ defect to the goal.

	These results indicate that the learned control policies are strong local optima, and may be close to global optima, for the control objectives considered here. Comparing controller trajectories with time-stepped reachable sets is thus a simple but illuminating method for assessing minimum-time control policies.

	\section{Summary and Discussion}
	\label{sec:summary}

	This work identifies several physical constraints on the controlled motion of active nematic defects. Finite-time reachable sets depend sensitively on the director field, homeotropic wall anchoring, channel geometry, and the spatial structure of the allowed activity patterns. Local activity patterns can steer defects in open domains, but they fail in junction geometries, where broken-symmetry director fields and wall anchoring create reachability barriers. Global patterns can overcome this barrier for some channels, while other targets become accessible only through pair creation and control of the newly created $+1/2$ defect. Reachability analysis therefore gives a direct way to characterize defect mobility and control landscapes in active nematics.

	Deep reinforcement learning can discover effective control sequences within these landscapes. Our RL controllers outperform both naive static patterns and rule-based controllers, which overshoot and can fail to prevent defect annihilation. By contrast, RL controllers transport the defect to the goal and stabilize it there with time-dependent activity patterns. The learned policies can also suppress or exploit the creation of new defects, leading to control strategies that would be difficult to design manually. In the maze geometry, controllers trained on simple junctions can be composed into a meta-controller for a larger test environment.

	The free energy analysis reveals a further physical effect. A guided $+1/2$ defect leaves persistent high-energy distortions in the director field, while a later $-1/2$ defect can partially erase this path history. Thus, controlled defect motion can write and erase memory in channel-confined active nematics. On the other hand, reachability-based optimality visualizations provide a complementary test of RL controller performance.

	Taken together, these results suggest that reinforcement learning can serve as a computational probe of active nematic physics. By searching over activity sequences, the learned policies expose reachability barriers, defect creation events, and history-dependent free energy landscapes that are difficult to infer from static patterns or hand-designed rules. The same methods should be adaptable to other soft and active matter control problems, especially for materials with complex dynamics where classical control methods are difficult to apply.

	The hybrid LBM simulations may also be extended to colloidal suspensions or multiphase flows. Experiments have already manipulated colloids in nematic fluids using optical tweezers and defect-assisted microrobots \cite{Musevic_2017, nematic_microrobots, phototunable_microrobots}. RL controllers could support automated colloidal assembly by reducing the need for manual steering. RL control of multiphase active flows could also aid microfluidic devices, labs-on-chips, micromixers, and microreactors, and may help improve our understanding of intracellular transport and jamming suppression in the cytoplasm \cite{Needleman_Dogic_2017, Kos_2019, manz_2021, Venkatesh_2022, PhysRevE.109.014606, Yang_Liu_2025}.

	Robustness and experimental feasibility remain open questions. Our simulations use ideal rectangular activity patterns, deterministic mean-field dynamics, and fixed boundary conditions, whereas experiments will include fluctuations, imperfect anchoring, camera and actuation latency, and optical activity profiles with blurred edges. Future numerical studies should test controllers under parameter shifts in elastic constants, viscosities, activity levels, anchoring strengths, delays, and pattern-edge smoothness.
	
	Because our RL controllers act in closed loop using observed fields, they should be able to compensate for moderate imperfections, especially if trained across various parameter ranges. Random fluctuations in the macroscopic Beris-Edwards fields may also improve generalization. Transfer learning could also be explored: a controller could first be trained on a simple geometry, then fine-tuned on progressively more complex domains. This may accelerate convergence on difficult control problems.

	Although we have focused on defect position control here, deep RL may also be applied to other tasks. Dynamic activity patterns could generate desired velocity profiles or director configurations in channel-confined active nematics. Separate RL controllers trained on different tasks can also be combined into higher-level planning algorithms, as demonstrated by our meta-controller, allowing multi-step active matter manipulations to be built from smaller trained components.

	Our work also fits into the broader context of PDE-constrained control and optimization. Liquid crystal hydrodynamics, including the Beris-Edwards equation, has attracted growing mathematical interest because it defines a challenging PDE system \cite{Ball_2017, Zarnescu_2021}. Surowiec and Walker \cite{Surowiec_Walker_2023} studied the optimal control of passive nematic simulations in a pioneering work, using boundary controls instead of activity patterns to drive director fields between equilibrium configurations. Their controller can manipulate preexisting defects and disclination lines using classical control theory, but unlike our RL framework, their method may not be suited to handling defect pair creation and annihilation.
	
	A detailed comparison between their method and ours could produce new insights. Classical adjoint-based optimization methods typically converge to local optima, so in highly nonconvex soft and active matter problems they may struggle to find globally optimal policies. Reinforcement learning does not provide rigorous optimality guarantees, but it can perform well in nonconvex spaces and produce robust controllers \cite{De_Lellis_2024}. Classical control and RL therefore have complementary strengths, and the best tool depends on the problem. Further comparisons between these approaches could connect developments in mathematics and physics, and may lead to a more unified theory of control in soft and active matter.

	\begin{acknowledgments}
		We wish to acknowledge Rui Zhang for his generous help regarding hybrid LBM implementation details. We also thank \v{Z}iga Kos and Miha Ravnik, whose 2D active nematics simulation tutorial \cite{Kos_Ravnik_code} was a source of inspiration for our code. Y. A. acknowledges support from JST FOREST Program (Grant No. JPMJFR222U), JST CREST (Grant No. JPMJCR23I2), and JST [Moonshot R\&D] (Grant No. JPMJMS256J). K. K. acknowledges support from JSPS KAKENHI (Grant Nos. JP23H00095 and JP25H01361).
	\end{acknowledgments}

	\appendix

	\section{The Beris-Edwards Equation and the Hybrid Lattice Boltzmann Method}
	\label{sec:app_be_eq}

	Let $Q$ be the symmetric, traceless tensor representation of the nematic director, varying in both space and time. The Beris-Edwards equation for the director field is \cite{Beris_Edwards_1994, Stewart_2004, Sonnet_Virga_2014, Zhang_LBM_2016}
	\begin{equation}
		\begin{split}
			\frac{\partial Q}{\partial t} + u \cdot \nabla Q - S &= \Gamma H ,
		\end{split}
	\end{equation}
	where $u$ is the velocity and $\Gamma$ is the molecular field strength.

	The advection term $S$ is
	\begin{equation}
		\begin{split}
			S = \left( \xi D + \Omega \right) \left( Q + \frac{I}{2} \right) &+ \left( Q + \frac{I}{2} \right) \left( \xi D - \Omega \right)\\
			&- 2 \xi \left( Q + \frac{I}{2} \right) \text{Tr} \left( Q W \right) ,
		\end{split}
	\end{equation}
	where the strain rate field is $D = \frac{1}{2} \left( W + W^T \right)$, the vorticity is $\Omega = \frac{1}{2} \left( W - W^T \right)$, the flow-aligning parameter is $\xi$, and the spatial derivatives of velocity are $W_{ij} = \partial_j u_i$.

	The molecular field is $H = -\frac{\delta f}{\delta Q}$, where the Landau-de Gennes (LdG) free energy density $f$ is
	\begin{equation}
		f = \frac{A}{2} \text{Tr} \left( Q^2 \right) + \frac{B}{3} \text{Tr} \left( Q^3 \right) + \frac{C}{4} \left( \text{Tr} \left( Q^2 \right) \right)^2 + \frac{L}{2} \left( \nabla Q \right)^2 ,
	\end{equation}
	where $A$, $B$, $C$, and $L$ are material-specific constants ($L$ is the Frank elastic constant). Taking the functional (Volterra) derivative, the explicit expression for $H$ is
	\begin{equation}
		H = -A Q - C Q \: \text{Tr} \left( Q^2\right) + L \, \nabla^2 Q .
	\end{equation}

	The Beris-Edwards equation is coupled to the damped Navier-Stokes equation
	\begin{equation}
		\rho \left( \frac{\partial}{\partial t} + u \cdot \nabla \right) u = \nabla \cdot \Pi - \mu u ,
	\end{equation}
	where $\rho$ is the density, $\mu$ is the damping coefficient, and the stress tensor is $\Pi = \Pi^a + \Pi^p$, where the active stress is $\Pi^a = -\alpha Q$ and the passive stress $\Pi^p$ is
	\begin{equation}
		\begin{split}
			\Pi^p = \; &2 \eta D - \rho T \: + \: \frac{L}{2} \left( \nabla Q \right)^2\\
			&+ 2 \xi \left( Q + \frac{I}{2} \right) \text{Tr} \left( Q H \right) - \xi H \left( Q + \frac{I}{2} \right)\\
			&- \xi \left( Q + \frac{I}{2} \right) H - L \, \nabla Q \odot \nabla Q + Q H - H Q .
		\end{split}
	\end{equation}
	Here $\eta$ is the dynamic viscosity, $T$ is the temperature, and $\left( \nabla Q \odot \nabla Q \right)_{ij} \equiv \partial_i Q_{\alpha \beta} \partial_j Q_{\alpha \beta}$.

	\begin{apstable}{tab:lbm-params}{Hybrid LBM (D2Q9) simulation parameters.}
		\begin{tabular*}{\columnwidth}{@{\extracolsep{\fill}} l r}
			Parameter name & Value (lattice units) \\
			\midrule
			Lattice dimensions & 420-by-420 \\
			LBM relaxation time & 1 \\
			FD steps per LBM step & 2 \\
			Timesteps per control step $\tau$ & $10^4$ \\
			Molecular field strength $\Gamma$ & 0.1 \\
			Flow-aligning parameter $\xi$ & 0.8 \\
			Damping coefficient $\mu$ & 0.01 \\
			Frank elastic constant $L$ & 0.1 \\
			LdG free energy coefficient $A$ & $-0.01667$ \\
			LdG free energy coefficient $B$ & $-0.35$ \\
			LdG free energy coefficient $C$ & 0.35 \\
			Activity level $\alpha_0$ & 0.0035 \\
		\end{tabular*}
	\end{apstable}

	We solve these equations using a GPU-optimized hybrid lattice Boltzmann method (LBM) with the D2Q9 velocity set. The implementation is written in C and CUDA \cite{yeomans_hlbm, Zhang_LBM_2016, Kos_Ravnik_code}, and all code is available at \cite{github_rl_defect_control_code}. The $Q$-tensor field is evolved using finite differences (FD). The simulation code supports arbitrary 2D axis-aligned microchannel geometries as well as periodic and open boundary conditions at channel outlets. Infinite homeotropic (wall-normal) boundary conditions are enforced along all obstacle walls.

	Simulation parameters are listed in lattice units in Table~\ref{tab:lbm-params}. These values correspond to a kinematic viscosity of $\nu \equiv \eta / \rho = 1 / 6$, a Reynolds number of $\text{Re} \equiv U \ell / \nu \approx 0.1$, and an active Ericksen number of $\text{Er}_a \equiv \alpha_0 \ell^2 / L \approx 130$ \cite{Giomi_Bowick_2014, chandler_active_ericksen}. The characteristic length is $\ell = 60$, equal to the microchannel width, and the characteristic velocity is $U = 3 \times 10^{-4}$, both in lattice units.

	\section{Algorithm Pseudocode}
	\label{sec:app_algos}

	Here we give pseudocode for reachable set estimation and our rule-based defect controller. Python implementations are available in our code repository \cite{github_rl_defect_control_code}.

	In Algorithm~\ref{alg:reachable-set-est}, activity patterns are randomly selected from the control pattern set and applied to the simulation. The simulation is reset when the episode length is reached, when the $+1/2$ defect is annihilated, or when the defect flows out of the domain under open boundary conditions. Reached $+1/2$ defect positions are stored in a list. After sampling, this list is passed to a concave-hull algorithm.
	
	\begin{figure}[t]
		\begin{apsalgorithm}{alg:reachable-set-est}{Reachable set estimation.}
			\Require Initial director and velocity fields $\phi_0$, empty list of reached positions $R$, number of sampling iterations $M$, episode length $N$, set of activity patterns $\{ P_i \}$, concave hull algorithm \Call{ConcaveHull}{}.
			\Ensure Concave hull of reached positions.
			\State Initialize simulation to $\phi_0$
			\State $k \gets 0$
			\For{each iteration $i \in ( 1, 2, \ldots, M )$}
			\If{$k = N$ or no $+1/2$ defect is found}
			\State Reset simulation to $\phi_0$
			\State $k \gets 0$
			\EndIf
			\State Select random activity pattern $P_i \in \{ P_i \}$
			\State Apply $P_i$ to simulation until next control step
			\State $\textbf{x}_d \gets$ Get $+1/2$ defect position from simulation
			\State Append $\textbf{x}_d$ to $R$
			\State $k \gets k + 1$
			\EndFor
			\State \Return \Call{ConcaveHull}{$R$}
		\end{apsalgorithm}
	\end{figure}
	
	\begin{figure}[t]
		\begin{apsalgorithm}{alg:rule-based-controller}{Rule-based defect controller.}
			\Require Goal position $\textbf{x}_g$, episode length $N$, set of local activity patterns $\{ P_i \}$.
			\Ensure Sequence of activity patterns $( P_j )$.
			\For {each control step $k \in (0, 1, \ldots, N - 1)$}
			\State $\textbf{x}_d \gets$ Get $+1/2$ defect position from simulation
			\For{each $P_i \in \{ P_i \}$}
			\State$\textbf{\^x}_{i}^{k + 1} \gets$ Predict defect position at next control
			\Statex \hspace{\algorithmicindent}\hspace{\algorithmicindent}\hspace{\algorithmicindent}step $k + 1$ under influence of $P_i$ on $\textbf{x}_d$
			\EndFor
			\State $j \gets \argmin_{i} \lVert \textbf{x}_g - \textbf{\^x}_{i}^{k + 1} \rVert$
			\State Apply $P_j$ to simulation until next control step
			\EndFor
		\end{apsalgorithm}
	\end{figure}

	Many open-source concave-hull generators are available. We use the \texttt{concave\_hull} Python library \cite{concave_hull}. If multiple $+1/2$ defects exist at a control step, one can either select the defect nearest a specified goal or append all such defects to the reached-position list. We use the first option.

	In the rule-based controller (Algorithm~\ref{alg:rule-based-controller}), local activity patterns are chosen deterministically based on the relative positions of the defect and goal. At each control step, the controller predicts the next $+1/2$ defect position for each candidate pattern. For the rectangular patterns used here, one end of the pattern is placed on the current defect position (see Fig.~\ref{fig:reach_sets_local}). The rule-based controller then naively predicts that the defect will move to the opposite end of the rectangle.

	The controller chooses the pattern whose predicted position is closest to the goal, and repeats this procedure at each step. The empty pattern, meaning no activity, is also included as a possible action. For the empty pattern, the rule-based controller assumes that the defect remains stationary until the next control step. This pattern is selected when the defect is already very close to the goal.

	Despite these strong assumptions, the rule-based controller performs comparably to the RL controller in the free geometry (Fig.~\ref{fig:perf_free_local8}). It fails, however, to steer defects through junction geometries. In less constrained environments, the rule-based controller may still perform reasonably well. If obstacles lie between the initial defect position and the goal, one should instead specify obstacle-free waypoints. Each adjacent pair of waypoints then defines a separate rule-based control problem.

	\section{RL Controller Implementation Details}
	\label{sec:app_rl_params}

	We train the RL controllers using proximal policy optimization (PPO), a policy-based reinforcement learning algorithm \cite{schulman2017ppo}. We use the PPO implementation in \texttt{stable-baselines3} \cite{stable-baselines3} and PyTorch \cite{pytorch} for neural network construction and training. This section summarizes the model architecture, reward shaping, and training times. Hyperparameters are listed in Table~\ref{tab:rl-hyperparams}.

	The model input contains normalized director and velocity fields in the $x$ and $y$ directions, plus a heatmap of defect positions. The resulting tensor has size 5-by-420-by-420, where 420 is the lattice side length. To construct the defect heatmap, we initialize a 420-by-420 matrix of zeros, add $+1$ at $+1/2$ defect positions and $-1$ at $-1/2$ defect positions, and then apply localized Gaussian smoothing. This produces a smooth field with well-behaved gradients.

	The first layers of the RL model alternate between convolutional layers and rectified linear unit (ReLU) layers, a standard architecture for grid-based inputs \cite{Lapan_2024}. The transformed data then enters a value network and a policy network, which share the same fully connected architecture. The final output dimension equals the action-space dimension and defines a probability distribution over actions. For discrete action spaces, corresponding to local control pattern sets, the controller selects the pattern with the highest probability. For multibinary action spaces, corresponding to global control pattern sets, all primitives with probabilities greater than 0.5 are applied simultaneously.

	We use reward shaping to accelerate convergence. The reward is proportional to the previous defect-to-goal distance minus the current distance, so the controller receives a positive reward when the defect approaches the goal and a negative reward when it moves away. Such difference-based rewards help prevent erroneous convergence to cyclic action sequences \cite{Ng_reward_shaping}. To encourage stabilization after arrival, we add a local bonus near the goal that linearly increases as the defect-to-goal distance decreases.

	A single simulation parameter controls whether the creation of new defect pairs is allowed. If creation is allowed, the controller switches to the $+1/2$ defect nearest the goal at each control step. If creation is forbidden, the episode is truncated and the simulation is reset as soon as a new defect pair forms, forcing the controller to avoid defect creation in order to increase episode length and total reward. Fig.~\ref{fig:perf_cross_straight}(a) shows a policy trained with defect creation forbidden, while Fig.~\ref{fig:perf_cross_up}(a) shows a policy trained with defect creation allowed. More generally, desired controller behaviors can be encouraged by adding reward terms, while unwanted behaviors can be discouraged by early episode truncation.

	All RL controllers were trained on a single NVIDIA GeForce RTX 5090 GPU with CUDA 12.9. The hybrid LBM simulation is the training bottleneck, so we implemented CUDA optimizations in C to improve memory access efficiency. With these optimizations, one full training episode takes roughly 30 seconds of wall time, corresponding to $20 \tau = 2 \times 10^5$ simulation timesteps, or 20 training/control steps.

	Policies for discrete action spaces converged after about $10^5$ training steps, or 40 hours of wall time. Models for multibinary action spaces took roughly twice as long because the number of possible activity patterns is much larger. We trained on one GPU to demonstrate the relatively low computational cost of the method, but because the hybrid LBM is highly parallelizable, training should scale efficiently to multi-GPU setups.

	\begin{apstable}{tab:rl-hyperparams}{RL controller hyperparameters.}
		\begin{tabular*}{\columnwidth}{@{\extracolsep{\fill}} l r}
			Parameter name & Value \\
			\midrule
			Control steps per model update & 512 \\
			Batch size & 256 \\
			Number of optimizer epochs per batch & 10 \\
			Learning rate & $2.5 \times 10^{-4}$ \\
			Entropy coefficient & $5 \times 10^{-3}$ \\
			Clipping parameter & 0.2 \\
			GAE lambda & 0.95 \\
			CNN filter dimensions per layer & [16, 32, 64, 64] \\
			CNN output dimension & 256 \\
			Policy network architecture & [256, 128] \\
			Value function architecture & [256, 128] \\
		\end{tabular*}
	\end{apstable}

	\section{T-Junction Controller Results}
	\label{sec:app_additional_results}

	\begin{figure*}[t]
	\centering
	\includegraphics[width=\textwidth]{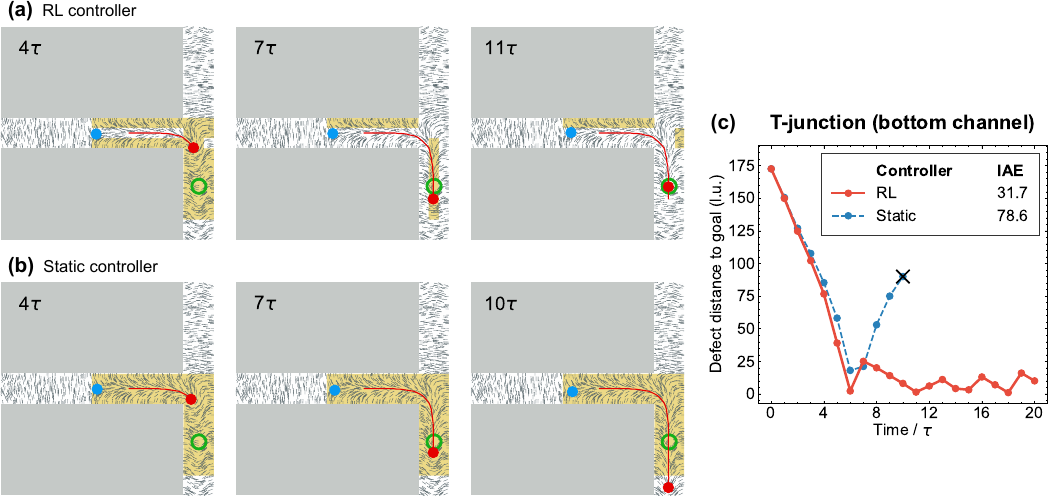}
	\caption{Defect trajectories on the T-junction with a bottom-channel goal, generated by (a) the RL controller and (b) the static controller. As in the cross junction, the defect reaches the goal under both controllers without defect creation or annihilation, after which only the RL controller stabilizes the defect near the goal. The error plot and IAE scores are shown in (c). The cross in the static-control curve indicates that the defect has flowed out of the domain.}
	\label{fig:perf_t_down}
	\end{figure*}

	\begin{figure*}[!t]
	\centering
	\includegraphics[width=\textwidth]{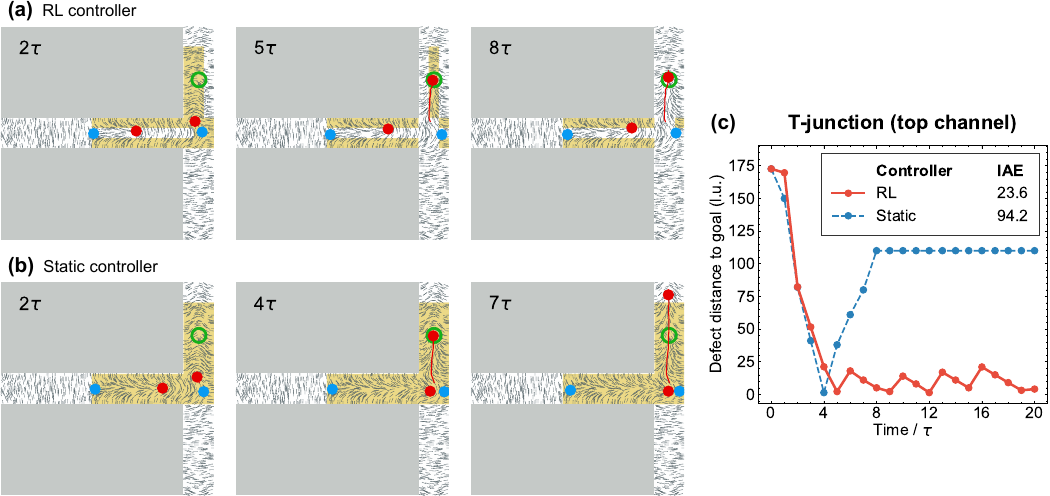}
	\caption{Defect trajectories on the T-junction with a top-channel goal, generated by (a) the RL controller and (b) the static controller. As seen for the cross junction, the initial $+1/2$ defect cannot enter the top channel intact. The RL controller learns to exploit defect creation in the junction, guide the new $+1/2$ defect to the goal, and trap it there. The error plot and IAE scores are displayed in (c).}
	\label{fig:perf_t_up}
	\end{figure*}

	Here we report additional RL and static controller results for the bottom and top channels of the T-junction (Figs.~\ref{fig:perf_t_down} and \ref{fig:perf_t_up}). These results provide a useful comparison with the cross-junction tasks in the main text. The T-junction removes one outlet from the cross junction geometry and uses open boundary conditions at the remaining outlets, but it retains the same basic ingredients that constrain defect motion: homeotropic walls, a broken-symmetry director field in the junction, and global activity primitives that can actuate both the channels and the intersection.

	The bottom-channel task tests whether the controller can guide the original defect through the T-junction and stabilize it near the goal. As in the cross-junction bottom-channel task, both static and RL control can move the original $+1/2$ defect into the target channel without creating or annihilating defect pairs. The difference appears after arrival. Under static control, the defect continues downstream and eventually leaves through the open outlet, which is marked by the cross in Fig.~\ref{fig:perf_t_down}(c). The RL controller instead modulates the activity pattern after arrival and holds the defect near the goal, lowering the time-averaged IAE.

	The top-channel task tests the same reachability limitation seen in the cross junction. Again, the original $+1/2$ defect cannot pass intact into the top channel of the T-junction. Instead, activity near the northeast corner creates a new defect pair, and the newly created $+1/2$ defect becomes the focus of control. The RL controller exploits this event deliberately: it guides the new defect to the goal and successfully holds it there.
	
	Thus, we see that the pair-creation strategy is common to both the cross junction and the T-junction. Evidently, open boundaries differ little from periodic ones in terms of their impact on both defect dynamics and learned control policies, at least for the geometries and tasks considered here. Still, these results demonstrate the flexibility of our controller development methodology in supporting different boundary conditions and microchannel configurations.

	\FloatBarrier

	\makeatletter
	\@namedef{b@apsrev42Control}{{}{}{{}}{{}}}
	\makeatother
	\nocite{apsrev42Control}
	\bibliographystyle{apsrev4-2}
	\bibliography{refs}
\end{document}